\documentclass[aps,preprint,showkeys]{revtex4}
\usepackage{amsfonts}
\usepackage{amsmath}
\usepackage{amssymb}
\usepackage{mathrsfs}
\usepackage{graphicx}
\usepackage{float}
\usepackage{color}
\usepackage{soul}
\usepackage{booktabs}

\usepackage{siunitx}
\usepackage{url}
\usepackage{CJKutf8}

\renewcommand {\thetable} {\arabic{table}}

\newcount\multicolscols

\newcommand{\appendixtablecaption}[2]{%
  \refstepcounter{table}\label{#1}%
  \noindent\textbf{Table~\thetable.}\enspace #2\par
  \vspace{0.15em}%
}

\begin{document}
\def\toprule{\noalign{\hrule height 0.8pt \vskip 2pt}}
\def\midrule{\noalign{\vskip 1pt \hrule height 0.5pt \vskip 2pt}}
\def\bottomrule{\noalign{\vskip 1pt \hrule height 0.8pt}}
\begin{CJK*}{UTF8}{gbsn}
\title{Revisiting the Lyth bound constraints on inflation from ACT DR6 results}
\thanks{Supported by the National Natural Science Foundation of China (12175212, 12275183, 12503005), the Sichuan Science and Technology Program (2026NSFSC0804), and the Fundamental Research Funds for the Central Universities (YJ202551).}
\author{Rui Yang (杨睿)}
\email{yangrui7@stu.scu.edu.cn}
\author{Jun Tao (陶军)}
\email{taojun@scu.edu.cn}
\author{Peng Wang (王鹏)}
\email{pengw@scu.edu.cn}
\author{Mian Zhu (朱勉)}
\email{zhumian@scu.edu.cn}
\thanks{(corresponding author)}
\affiliation{College of Physics, Sichuan University, Chengdu 610065, China}

\begin{abstract}
The Lyth bound states that the field excursion of the inflaton must be sub-Planckian, thereby imposing an upper bound on the amplitude of the tensor power spectrum in inflationary scenarios. This bound is conventionally derived under the assumption of a scale-invariant curvature power spectrum, i.e., $n_s = 1$. However, astrophysical observations confirm a red-tilted spectrum with $n_s < 1$. In light of recent results from the Atacama Cosmology Telescope (ACT) DR6, we revisit these constraints using the newly inferred scalar spectral index, $n_s \simeq 0.9743$. Incorporating the ACT data yields a different upper bound on the tensor-to-scalar ratio $r$ and clarifies whether non-canonical inflationary scenarios can remain compatible with both CMB observations and a sub-Planckian field excursion. Our result highlights the importance of combining theoretical considerations of the Lyth bound with up-to-date astrophysical data.
\end{abstract}

\maketitle
\end{CJK*}
\flushbottom

\section{Introduction}
\label{secintro}

Inflation is the leading paradigm for the primordial universe \cite{Guth:1980zm,Linde:1981mu,Albrecht:1982wi,Starobinsky:1980te,Mukhanov:1981xt,Baumann:2009ds,Martin:2013tda}. Nevertheless, inflation faces conceptual challenges. A central challenge is the so-called Lyth bound \cite{Lyth:1996im}, which calls into question the consistency of an effective field theory description of inflation. The Lyth bound states that the field excursion $\Delta \phi$ is related to the tensor-to-scalar ratio $r$ and the e-folding number $N_{\ast}$ via
\begin{equation}
\label{eq:Lythorigin}
    \Delta \phi/M_p \simeq \left( \frac{r}{0.002} \right)^{\frac{1}{2}} \left( \frac{N_{\ast}}{60} \right) ~,
\end{equation}
in canonical inflation. As a result, the field excursion over the entire inflationary period can exceed the Planck mass when $r \geq 0.002$. This implies that the effective description is sensitive to higher-dimensional operators unless the large field range is protected by a symmetry or by a suitable ultraviolet completion. Extensive literature has investigated the implications of the Lyth bound and possible ways to circumvent it \cite{Efstathiou:2005tq,Alishahiha:2004eh,Baumann:2006cd,Antusch:2014cpa,Stein:2016jja,Planck:2018jri,Bravo:2020wdr,BICEP:2021xfz,Cai:2021yvq,Paoletti:2022anb,AtacamaCosmologyTelescope:2025nti,Furuta:2025tjt,Akama:2026haf,Xue:2008mk,Ashoorioon:2014nta,Ashoorioon:2013eia,Li:2015sja,Zhang:2007bi,Hartong:2006rt,Gu:2023mmd,Gu:2026ajw,Zhang:2019xek,Li:2024fxy}. See, e.g., \cite{Huang:2015xda} for a review.

Conventionally, the Lyth bound is evaluated by assuming a scale-invariant curvature power spectrum $P_{\zeta}$, i.e., a scalar spectral index $n_s = 1$. However, observations suggest a slight red tilt in $P_{\zeta}$. For example, the Planck 2018 results suggest \cite{Planck:2018vyg}
\begin{equation}
    n_s = 0.9649 \pm 0.0042 ~.
\end{equation}
When the red tilt is taken into consideration, the constraints imposed by the Lyth bound are modified \cite{Efstathiou:2005tq,Easther:2006qu}. Notably, Refs. \cite{Garcia-Bellido:2014eva,Garcia-Bellido:2014wfa} proposed that this red tilt strengthens the Lyth bound constraint. Therefore, it is essential to revisit the Lyth bound constraint by accounting for the red tilt. This task is especially timely given the recent Atacama Cosmology Telescope (ACT) result, as the ACT DR6 data imply a scalar spectral index slightly different from that reported by the Planck Collaboration
\cite{AtacamaCosmologyTelescope:2025blo}
\begin{equation}
\label{eq:nsACT}
    n_s = 0.9743 \pm 0.0034 ~,
\end{equation}
and has generated immediate interest in the cosmology community \cite{Kallosh:2025rni,Dioguardi:2025vci,Ferreira:2025lrd,McDonough:2025lzo,Zhu:2025twm,Gao:2025viy,Yi:2025dms,He:2025bli,ACT:2023dou,Pang:2024qyh,Gao:2025onc,Liu:2025qca,Peng:2025bws,Li:2025ops,Peng:2025vda,Peng:2025tqt,Wu:2025vfs,Fu:2025ciy,Feng:2026pzs,Zhang:2025zgo,Yang:2026flt,Choi:2025qot,Odintsov:2025eiv,Kim:2025dyi,Yogesh:2025wak,Mohammadi:2025gbu,Addazi:2025agg,Addazi:2025qra,Zhang:2026ivx,Wang:2025cpp,Sabogal:2026qvy,Qiu:2025iqm,Oikonomou:2026qkj,Odintsov:2026doe}. Hence, we are motivated to investigate the updated Lyth bound constraints using the new ACT DR6 data.

Because canonical single-field slow-roll inflation is severely constrained by the Lyth bound, we consider inflationary scenarios with non-canonical scalar-field kinetic terms, which are constructed to evade the Lyth bound challenge \cite{Armendariz-Picon:1999hyi,Garriga:1999vw,Cheung:2007st,Chen:2006nt,Gao:2011qe,Meng:2004ap,Fu:2019vqc,Kuang:2016edj}. Notably, the Lyth bound constraints on these non-canonical scenarios have been investigated almost exclusively under the assumption $n_s = 1$, and can therefore be improved by incorporating the ACT results, which is the focus of this manuscript. Specifically, we investigate two possible non-canonical couplings: one with a kinetic-inflaton coupling \cite{Wetterich:2013wza,Wetterich:2014eaa,Hossain:2014xha,Hossain:2014coa,Hossain:2014ova}, and another with a modified kinetic term $K(X)$ \cite{Mukhanov:2005bu,Unnikrishnan:2008ki,Unnikrishnan:2012zu,Li:2012vta,Pareek:2021lxz,Lola:2020lvk,Zhu:2014wfa,Liu:2024xkz}. We find that the combination of the sub-Planckian field excursion required by the Lyth bound and the red-tilted curvature spectrum suggested by ACT results can impose strengthened constraints on inflationary scenarios.

The manuscript is organized as follows. We introduce the concept of the Lyth bound pedagogically in Sec. \ref{sec:LB}, and illustrate how a red-tilted curvature spectrum can strengthen this bound, using a kinetic-inflaton coupling model as an example. We then investigate inflationary scenarios with modified kinetic terms in Sec. \ref{sec:modifyk}, and show that a certain class of models consistent with the PLANCK 2018 + Lyth bound joint constraints is excluded by the ACT + Lyth bound joint analysis. We finally conclude in Sec. \ref{sec:conclusion}.

\section{A pedagogical introduction to Lyth bound}
\label{sec:LB}

\subsection{The original Lyth bound}
In a spatially flat Friedmann--Lemaître--Robertson--Walker (FLRW) universe
\begin{equation}
  ds^2 = a^2(\tau) \left[ -d\tau^2 + \delta_{ij}dx^i dx^j \right] ~,
\end{equation}
\begin{equation}
    \frac{1}{M_p} \left|\frac{d\phi}{dN} \right| = \sqrt{2\epsilon} ~.
\end{equation}
Moreover, the consistency relation implies $r = 16\epsilon$, so that 
\begin{equation}
    \frac{1}{M_p} \left| \frac{d\phi}{dN} \right| \simeq \sqrt{\frac{r}{8}}.
\end{equation}
Thus, the canonical field excursion during inflation is
\begin{equation}
    \frac{\Delta\phi}{M_p} \simeq \int_{N_{\rm bg}}^{N_{\rm end}} dN     \sqrt{\frac{r(N)}{8}} ~,
\label{eq:field_range}
\end{equation}
where $N_{\rm bg}$ and $N_{\rm end}$ denote the number of e-folds at the beginning and end of the time interval under consideration, respectively. For later convenience, let us introduce $N_\star \equiv N_{\rm end}-N_{\rm bg}$. We require the single-field description to remain valid at least from the CMB scale to the end of inflation. To solve the flatness and horizon problems, this interval is conventionally estimated to be $N_{\ast} \simeq 60$. Assuming a constant $r$ over different values of $N$, the integral immediately yields the Lyth bound constraint in Eq. \eqref{eq:Lythorigin}. 

\subsection{An example: Lyth bound with non-canonical kinetic terms}
To demonstrate how the inclusion of non-canonical kinetic terms may modify the Lyth bound constraint, we begin with the following non-canonical action.
\begin{equation}
\label{eq:actionI}
    \mathcal{S} = \int d^4x \sqrt{-g} \left[ \frac{M_p^2}{2} R -\frac{1}{2} m^2(\phi) \partial^\mu \phi \partial_\mu\phi -V(\phi) \right] ~,
\end{equation}
and use the following choices \cite{Hossain:2014xha}
\begin{align}
m^2(\phi) = \left(\frac{\alpha^2-\tilde{\alpha}^2}{\tilde{\alpha}^2}\right) \frac{1}{1+\beta^2 e^{\alpha\phi/M_p}}+1 ~,~ V(\phi) = M_p^4 e^{-\alpha\phi/M_p} ~.
\label{eq:potential}
\end{align}
Here, the coupling $m^2(\phi)$ can reduce the field excursion relative to the canonical case for a given geometric evolution. Although we adopt the form of $m^2(\phi)$ following \cite{Hossain:2014xha}, any choice satisfying $m(\phi) \gg 1$ can, in principle, produce the same effect. In addition, the steep exponential potential \cite{Geng:2015fla} can render the field coordinate effectively invariant after canonical normalization and is therefore widely adopted in quintessential inflation.
The associated Friedmann equations are
\begin{align}
3M_p^2 H^2 &= \frac12 m^2(\phi)\dot{\phi}^2+V(\phi) ~, \\
2M_p^2\dot H &= -m^2(\phi)\dot{\phi}^2 ~.
\end{align}
The field velocity in this case is
\begin{equation}
\frac{1}{M_p} \left| \frac{d\phi}{dN} \right| = \frac{\sqrt{2\epsilon}}{m(\phi)} ~.
\end{equation}
Compared with Eq. \eqref{eq:field_range}, we find that $\Delta \phi$ can be modified by the function $m(\phi)$. Thus, an appropriately chosen $m(\phi)$ can reduce the value of $\Delta \phi$, thereby allowing a smaller field excursion for a fixed $r$ \cite{Hossain:2014xha}. Although this appears to circumvent the Lyth bound for the original field $\phi$, such a conclusion is premature, as demonstrated in the following section.

\subsection{Coordinate excursion versus canonical excursion}
\label{sec:canonical}
This apparent evasion, however, can be an artifact of field parametrization. Under a field redefinition $\phi \to f(\phi)$, the excursion changes, whereas the physics remains invariant. A commonly adopted approach in discussions of the Lyth bound is to measure the distance using the canonically normalized field \cite{Baumann:2011ws}. In the model in Eq. \eqref{eq:actionI}, this can be done naturally by introducing the canonically normalized field
\begin{equation}
    d\chi \equiv m(\phi) d\phi ~.
\end{equation}
However, after transforming to the canonical scalar field, the field range is simply
\begin{equation}
\label{eq:deltachi}
    \frac{\Delta \chi}{M_p} = \int^{N_{\ast}} dN \sqrt{r(N)/8} ~.
\end{equation}
Thus, if the field excursion is interpreted as the canonical field excursion, the kinetic coupling to the inflaton does not enable the model to evade the constraints imposed by the Lyth bound.

Specifically, the excursion of $\phi$ in the model of Eq. \eqref{eq:actionI} is 
\begin{equation}
\label{eq:deltaphi1}
\frac{1}{M_p} \left| \frac{d\phi}{dN} \right| = \frac{\sqrt{2\epsilon}}{m(\phi)} = \frac{\alpha}{m^2(\phi)}= \frac{2\epsilon}{\alpha} = \frac{r}{8\alpha} ~.
\end{equation}
where we used the expression for the slow-roll parameter $\epsilon \equiv -\dot{H}/H^2$ 
\begin{equation}
m(\phi(N_p))^2 \simeq \frac{8\alpha^2 }{r} ~,~ \epsilon \simeq \frac{M_p^2}{2m^2(\phi)} \left(\frac{V'(\phi)}{V(\phi)}\right)^2 = \frac{\alpha^2}{2m^2(\phi)} ~,
\end{equation}
and it is immediately apparent that a large $\alpha$ can significantly suppress $\Delta \phi$.

Moreover, we need to account for the effect of $n_s \neq 1$ on $\Delta \chi$. Again, the consistency relation gives
\begin{equation}
    n_t \simeq -\frac{r}{8} ~,~ \delta \equiv 1-n_s ~,
\end{equation}
thus 
\begin{equation}
    \frac{d\ln r}{dN} \simeq n_t-(n_s-1) = \delta-\frac{r}{8} ~.
\end{equation}
Given the observed small red tilt, we adopt the approximation $r(N) \simeq r_{\ast} e^{\delta N}$, where $r_{\ast}$ is the tensor-to-scalar ratio at the ACT pivot scale, $k_{\ast} = 0.05 Mpc^{-1}$. This leads to
\begin{equation}
    \Delta \chi/M_p \simeq \left( \frac{r_{\ast}}{8} \right)^{\frac{1}{2}} \frac{2}{\delta} \left( e^{\delta N_{\ast}/2} - 1 \right) ~,
\end{equation}
which reduces to $\Delta \chi/M_p = N_{\ast} (r_{\ast}/8)^{1/2}$ in the limit $\delta \to 0$. Notably, the dependence of $\chi$ on $\alpha$ is now implicit. In addition, the inequality $\Delta \chi /M_p \ll 1$ implies
\begin{equation}
\label{eq:rupper}
    r_{\ast} \ll \frac{2\delta^2}{\left( e^{\delta N_{\ast}/2} - 1 \right)^2} ~.
\end{equation}
Therefore, the upper bound on $r_{\ast}$ cannot be adjusted by varying $\alpha$ once $\delta$ and $N_{\ast}$ are fixed.

\subsection{Improved Lyth bound constraints with ACT}
\label{sec:ACTmodel1}
Here, we do not intend to assess whether $\Delta \phi$ and $\Delta \chi$ are more appropriate for considerations of the Lyth bound. Instead, we show in this section that precise measurements of $n_s$ and $r$ are essential for constraining the Lyth bound and that the model can be constrained through a joint analysis of the Lyth bound and ACT observations. The model in Eq.~\eqref{eq:actionI} is designed to suppress the coordinate excursion and thereby evade the usual Lyth bound constraint. Thus, the Lyth bound enters the scenario at the level of model construction. Nevertheless, once the action is specified based on the Lyth bound consideration, its predictions for $(n_s,r)$ cannot be consistent with ACT observations, as shown below. For the canonical field excursion case, we derive the upper bound on $r$ from Eq. \eqref{eq:rupper} for three cases: $n_s = 1$, $n_s = 0.9649$ (Planck 2018), and $n_s = 0.9743$ (ACT DR6). The results are summarized in Table  \ref{tab:rbd_planck_act_comparison}. We find that the Lyth bound becomes more stringent as the deviation of $n_s$ from unity increases. The Planck 2018 results impose the strictest upper bound, $r < 7.1 \times 10^{-4}$ for $N_{\ast} = 60$, whereas the ACT results slightly alleviate this constraint, allowing $r \lesssim 9.7 \times 10^{-4}$, because the suggested value of $n_s$ is slightly larger than that from the Planck result.

\begin{table}[htbp]
\centering
\begin{tabular}{cccc}
\toprule
$N_*$
& \begin{tabular}{c}
$\delta \to 0$ \\
$n_s=1$
\end{tabular}
& \begin{tabular}{c}
Planck 2018 \\
$n_s=0.9649$
\end{tabular}
& \begin{tabular}{c}
ACT DR6 \\
$n_s=0.9743$
\end{tabular}
\\
\midrule
50
& $3.20\times 10^{-3}$
& $1.25\times 10^{-3}$
& $1.63\times 10^{-3}$
\\
60
& $2.22\times 10^{-3}$
& $7.07\times 10^{-4}$
& $9.78\times 10^{-4}$
\\
70
& $1.63\times 10^{-3}$
& $4.22\times 10^{-4}$
& $6.21\times 10^{-4}$
\\
80
& $1.25\times 10^{-3}$
& $2.61\times 10^{-4}$
& $4.10\times 10^{-4}$
\\
\bottomrule
\end{tabular}
\caption{Upper limit on $r$ from the Lyth bound for different assumed values of $n_s$.}
\label{tab:rbd_planck_act_comparison}
\end{table}

\begin{figure*}[h]
    \centering
    \includegraphics[width=0.485\linewidth]{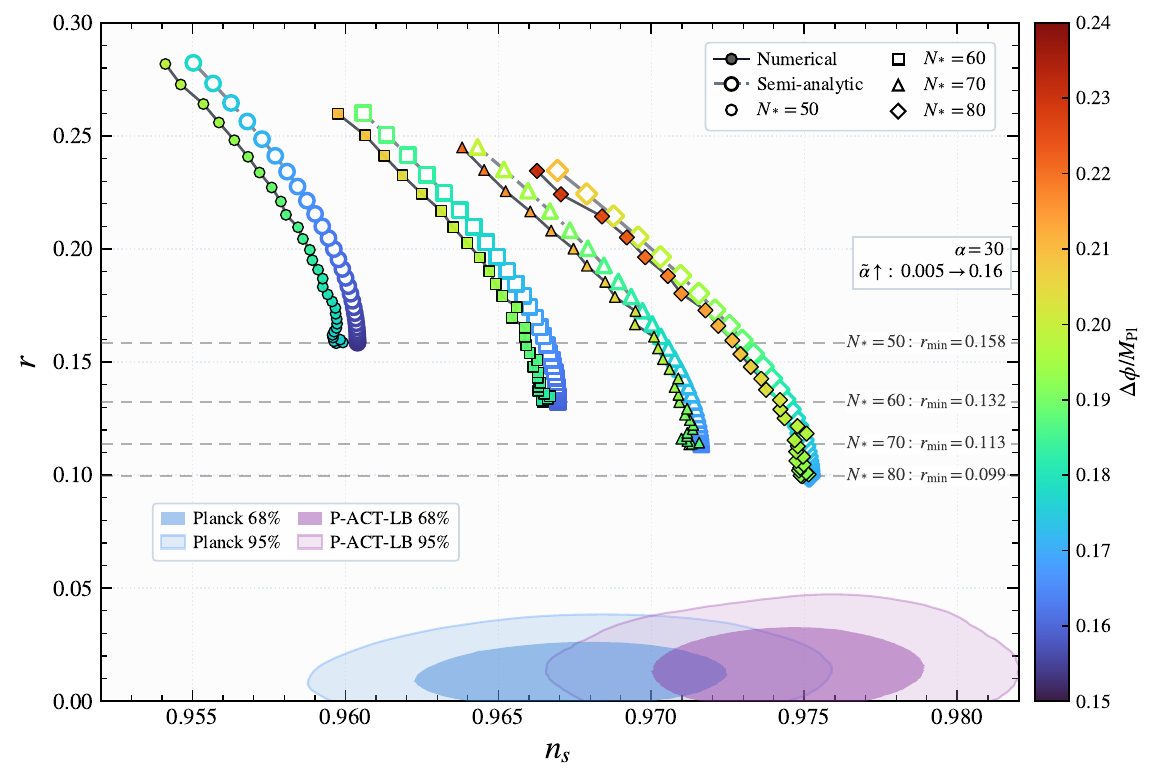}
    \includegraphics[width=0.485\linewidth]{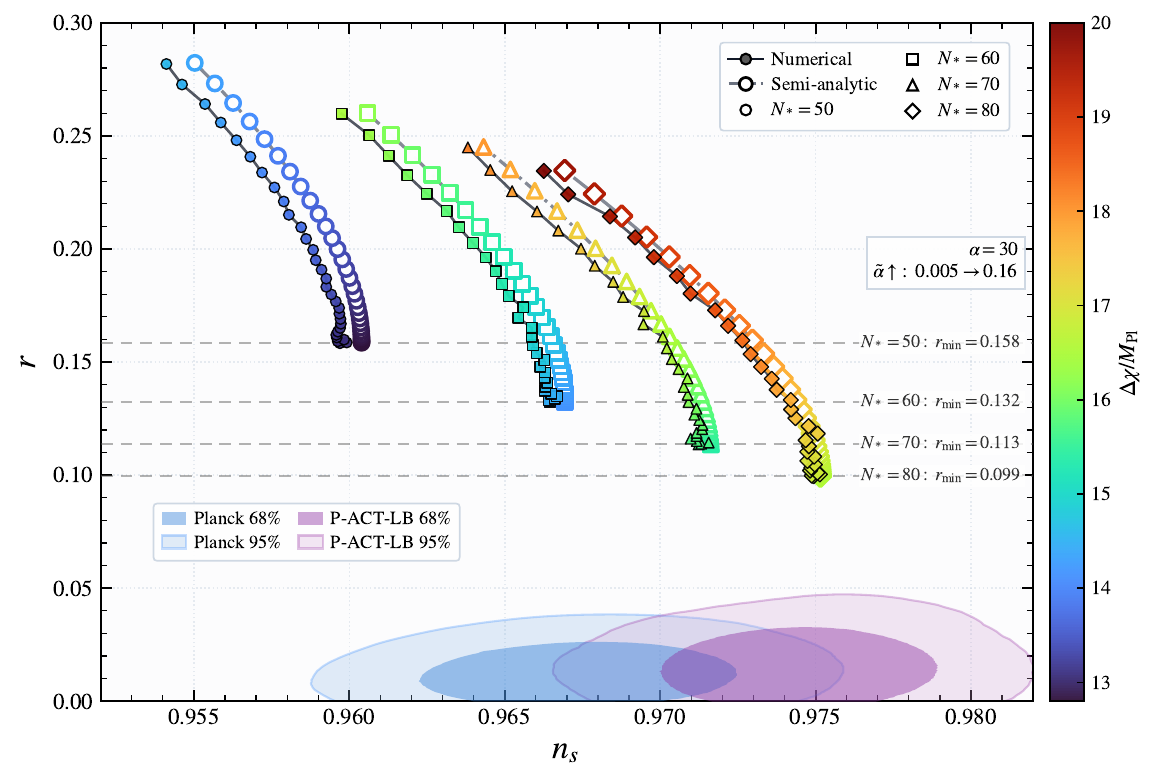}
    \caption{
    The $(n_s,r)$ plot is shown for model parameter $\alpha=30$ and $N_*=50,60,70,80$. The model parameter is chosen such that $A_s = 2.1 \times 10^{-9}$ is fixed. The numerical and semi-analytic $(n_s,r)$ values are listed in Tables~\ref{tab:A1}--\ref{tab:A4} and \ref{tab:A5}--\ref{tab:A8} of Appendix \ref{Model I Data}. The left and right panels show the coordinate and canonical field excursions, $\Delta \phi$ and $\Delta\chi$, respectively, indicating that the canonical field excursion is far beyond the Planck scale.} 
    \label{fig:nsrI}
\end{figure*}

We now consider the original field excursion $\Delta \phi$. Although the model permits a small $\Delta \phi$ for $\alpha \gg 1$, an internal consistency relation between $n_s$ and $r$ could render the model problematic. In the slow-roll approximation, 
\begin{equation}
    A_s = \frac{H^2}{8\pi^2M_p^2\epsilon} ~,~ n_s-1 = -2\epsilon - \eta ~,~ r = 16\epsilon ~,
\end{equation}
where $\eta \equiv \dot{\epsilon}/H\epsilon$ is the second slow-roll parameter. Using the Friedmann equations and the slow-roll condition, we obtain
\begin{equation}
     \eta(r) \simeq \frac{(8\alpha^2-r)(r-8\tilde{\alpha}^2)}{64(\alpha^2-\tilde{\alpha}^2)} ~.
\end{equation}
Expressing $n_s$ in terms of $\eta$ yields the following consistency relation 
\begin{equation}
r(n_s) = 8\left[ \alpha^2 - \sqrt{ (\alpha^2-\tilde{\alpha}^2) \bigl(\alpha^2-(1-n_s)\bigr) } \right] ~,
\label{eq:rnsconstraint}
\end{equation}
In the large-$\alpha$ limit, which is required for a small $\Delta \phi$, Eq. \eqref{eq:rnsconstraint} reduces to 
\begin{equation}
    r \simeq 4 ( \tilde{\alpha}^2 + \delta ) \geq 4\delta ~. 
\end{equation}
The ACT results imply $\delta = 0.0257$ and therefore $r \geq 0.103$, which far exceeds the upper limit set by the ACT observations. We perform both a semi-analytic investigation using the slow-roll approximation and a full numerical evaluation, as summarized in Fig. \ref{fig:nsrI}. The small differences between the semi-analytic and numerical results mainly arise from the slow-roll approximation used in the semi-analytic calculation, whereas the numerical calculation retains the full time evolution of the background and perturbations. The lower bound on $r$ from Eq. \eqref{eq:rnsconstraint} indeed prevents the model from being consistent with the ACT observations. The numerical and semi-analytic $(n_s,r)$ values plotted in both panels of Fig. \ref{fig:nsrI} are listed in Tables~\ref{tab:A1}--\ref{tab:A4} and \ref{tab:A5}--\ref{tab:A8} of Appendix \ref{Model I Data}. 

The preceding discussion shows that the deviation of $n_s$ from unity can significantly affect the constraints imposed by the Lyth bound. In particular, a joint analysis of the Lyth bound and the observed $(n_s, r)$ values from CMB data can substantially strengthen constraints on inflationary models. Specifically, we show that the model \eqref{eq:actionI}, which was designed to evade the theoretical constraint from the Lyth bound, fails to pass the ACT test. 

\section{Inflation with modified kinetic terms}
\label{sec:modifyk}

We now consider the inflationary scenario with modified kinetic terms. We adopt the following Lagrangian 
\begin{equation}
\label{eq:PX}
    P(X,\phi) = \mathcal K(X)-V(\phi) ~,~ X\equiv -\frac12 g^{\mu\nu}\partial_\mu\phi\,\partial_\nu\phi ~.
\end{equation}
In a homogeneous background, $X=\dot\phi^2/2$. The tensor-to-scalar ratio $r$ can be suppressed by using a power-law parametrization of $K(X)$ 
\begin{equation}
\label{eq:KX}
    \mathcal K(X) = M^4 \left(\frac{X}{M^4}\right)^\gamma ~,~ \gamma\geq 1 ~,
\end{equation}
because the sound speed of curvature fluctuations is modified to 
\begin{equation}
\label{eq:cs2}
    c_s^2 \equiv \frac{P_{,X}}{P_{,X}+2XP_{,XX}} = \frac{1}{2\gamma-1} < 1 ~,
\end{equation}
and, as a result, the consistency relation $r \simeq 16 c_s \epsilon$ implies a smaller $r$ for the same slow-roll parameter.
The power-law parametrization in Eq. \eqref{eq:KX} is a representative model for the $k$-inflation scenario, as it keeps the sound speed constant. In particular, since $r \propto c_s$, reducing $c_s$ can help suppress $r$, motivating phenomenological interest in investigating such models.

Let us explain why the model in Eq. \eqref{eq:KX} is expected to significantly modify the field excursion. For illustrative purposes, we use an exponential potential \cite{Geng:2015fla}
\begin{equation}
    V(\phi) = V_0 \exp \left[ -\lambda\left(\frac{\phi}{M_p}\right)^n \right] ~,
\end{equation}
where $M$ is a parameter with dimensions of mass. The slow-roll condition can be satisfied for $\lambda>0$ and $n > 0$, with the slow-roll parameter given by 
\begin{equation}
    \epsilon \equiv -\frac{\dot H}{H^2} = \frac{XP_{,X}}{M_p^2H^2} = \frac{\gamma X^\gamma}
    {M_p^2H^2 (M^4) ^{\gamma-1}} ~.
\end{equation}
Without loss of generality, we take $\phi>0$; the field excursion is
\begin{equation}
    \label{eq:DeltaN}
    \frac{1}{M_p}\frac{d\phi}{dN} = \sqrt{2}
    \left(\frac{\epsilon}{\gamma}\right)^{1/(2\gamma)}
    \left(\frac{M}{M_p}\right)^{1-1/\gamma}
    \left(\frac{H}{M}\right)^{1/\gamma-1} ~.
\end{equation}
It is immediately apparent that the tensor-to-scalar ratio $r$ cannot, by itself, determine the field excursion. In particular, in the limit $\gamma \gg 1$, the field excursion reduces to 
\begin{equation}
    \frac{1}{M_p} \frac{d\phi}{dN} \simeq  \sqrt{2} \frac{M^2}{M_pH} ~,
\end{equation}
which is completely decoupled from $r$. Thus, these models are expected to bypass the Lyth bound constraints through a substantial modification of the $\Delta \phi$-$r$ relation \cite{Unnikrishnan:2012zu,Li:2012vta}. 

So far, the discussion has been based on the original field excursion, while a canonically normalized field cannot be introduced in the same way as in Sec. \ref{sec:canonical}. The model \eqref{eq:KX} is genuinely nonlinear in $X$, and there is, in general, no global field redefinition $\phi \rightarrow \chi(\phi)$ that transforms the full Lagrangian into a canonical form. Therefore, we use the original field excursion $\Delta \phi$ as a reference to analyze the Lyth bound constraint.

\begin{figure}[ht]
    \centering
    \includegraphics[width=0.48\linewidth]{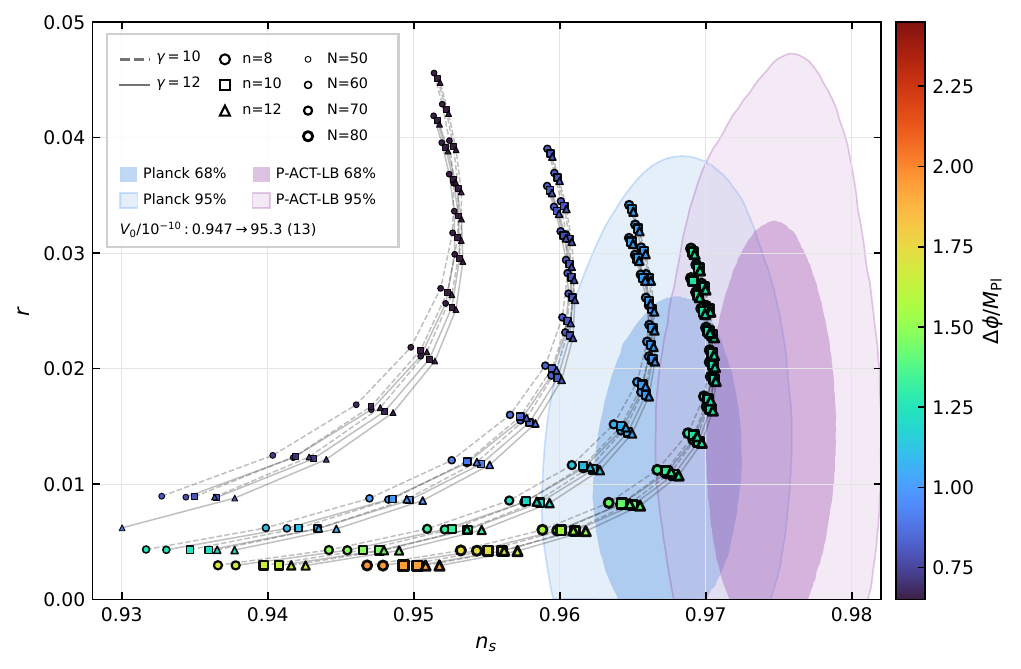}
    \includegraphics[width=0.45\linewidth]{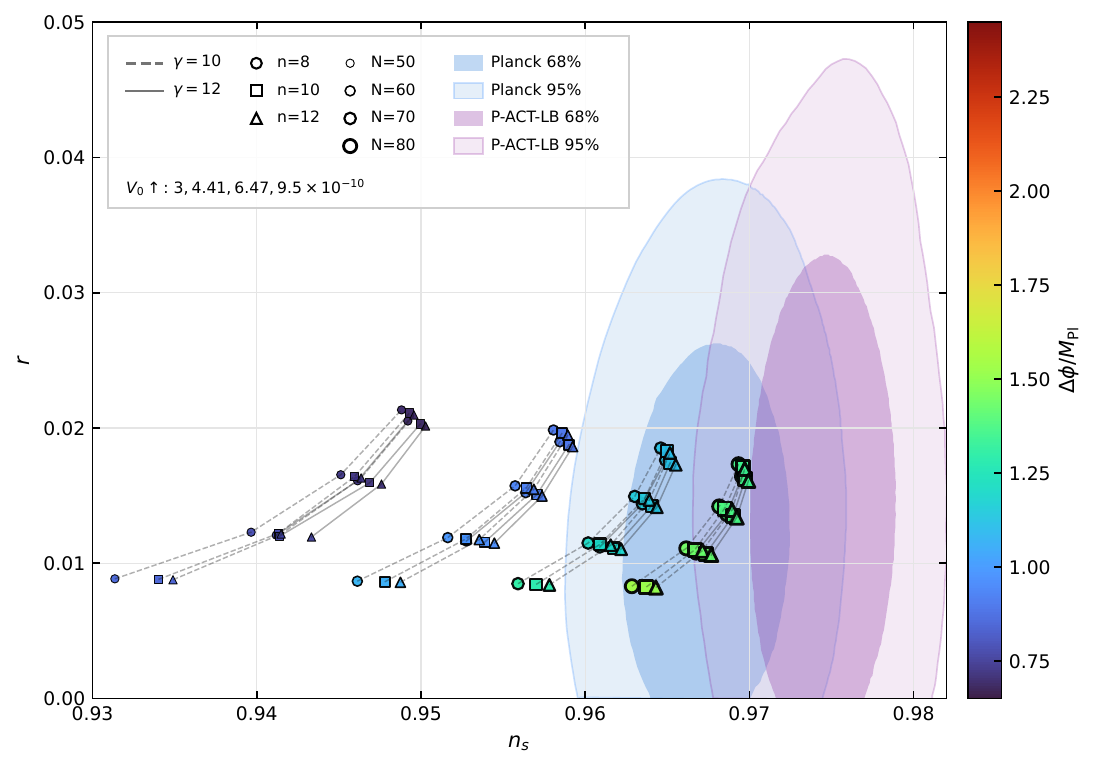}
    \caption{Left panel: semi-analytic results for $(n_s,r)$ and $\Delta \phi$ based on Eq. \eqref{eq:DeltaN}, whose validation assumes the slow-roll approximation. Right panel: numerical results obtained by directly evolving $\zeta$ for comparison. Again, the model parameter is chosen such that $A_s = 2.1 \times 10^{-9}$ is fixed. The semi-analytic results in the left panel are listed in Tables~\ref{tab:B25}--\ref{tab:B48} of Appendix \ref{app:model2}; the numerical results in the right panel are listed in Tables~\ref{tab:B1}--\ref{tab:B24}. The color bar denotes the original field excursion $\Delta\phi/M_p$.}
    \label{fig:model2res}
\end{figure}

To examine whether the model in Eq. \eqref{eq:PX} can indeed bypass the Lyth bound constraint while remaining consistent with ACT observations, we evaluate $(n_s,r)$ and $\Delta \phi$ for different model parameters, both semi-analytically under the slow-roll approximation and numerically. Tables~\ref{tab:B25}--\ref{tab:B48} of Appendix \ref{app:model2} list the semi-analytic results shown in the left panel of Fig. \ref{fig:model2res}, while Tables~\ref{tab:B1}--\ref{tab:B24} list the numerical results shown in the right panel. 
The origin of the small difference between the semi-analytical and numerical results is the same as that discussed for Model I.
The results show that ACT data, compared with Planck data, actually strengthen the Lyth bound constraint, in contrast to the case study in Sec. \ref{sec:ACTmodel1}. In particular, there exists a parameter space that fits the Planck data well while being marginally consistent with the Lyth bound, i.e., $\Delta \phi/M_p \simeq 1$. However, the field excursion $\Delta \phi/M_p$ must exceed unity to be consistent with ACT observations in this model. Thus, the update from ACT measurements puts models that were originally consistent with Planck under tension.

One may wonder whether an extreme choice of parameters can restore the viability of the model. To discuss this possibility, we analyze the effects of the model parameters on $n_s$, $r$, and $\Delta \phi$. First, the e-folding number $N$ has a critical impact on the values of $n_s$ and $r$: increasing $N$ enhances $n_s$ while keeping $r$ almost constant, at the cost of increasing $\Delta \phi$. When $\gamma$ is of order $10$, $N$ must approach $80$ to satisfy the ACT observations. However, in this case, $\Delta \phi / M_p$ already exceeds 1.5; thus, changing $N$ does not help substantially. 

\begin{figure}[ht]
    \centering
    \includegraphics[width=0.45\linewidth]{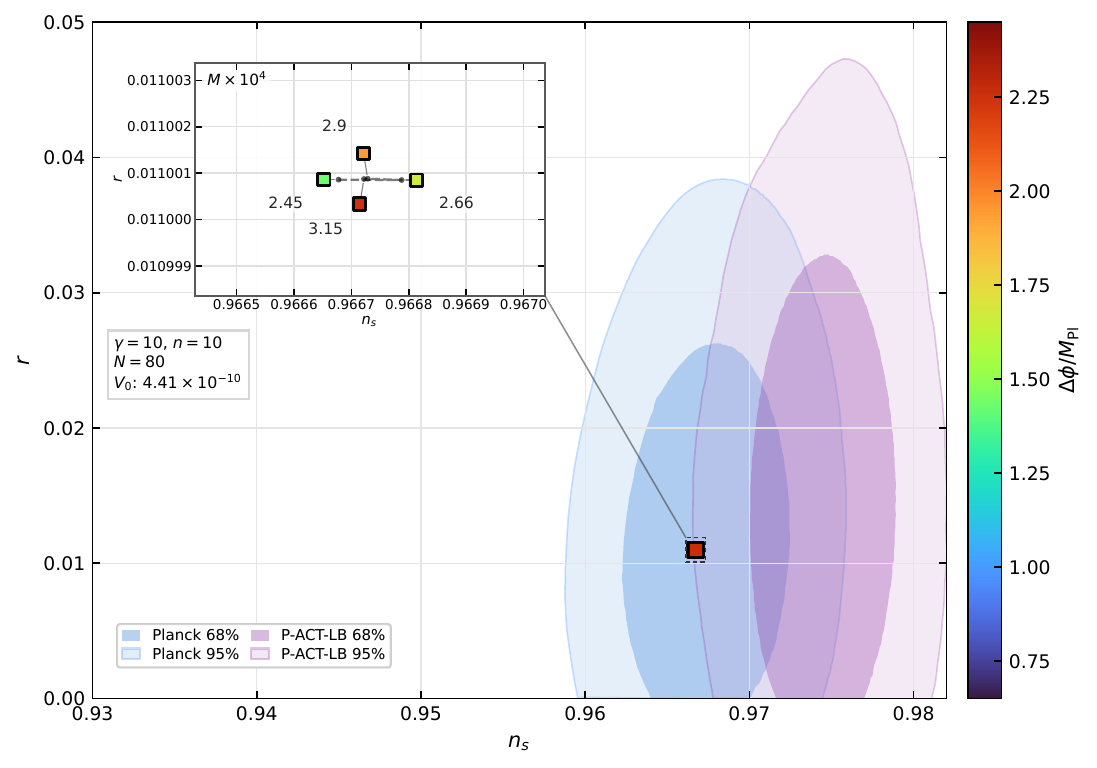}
    \includegraphics[width=0.45\linewidth]{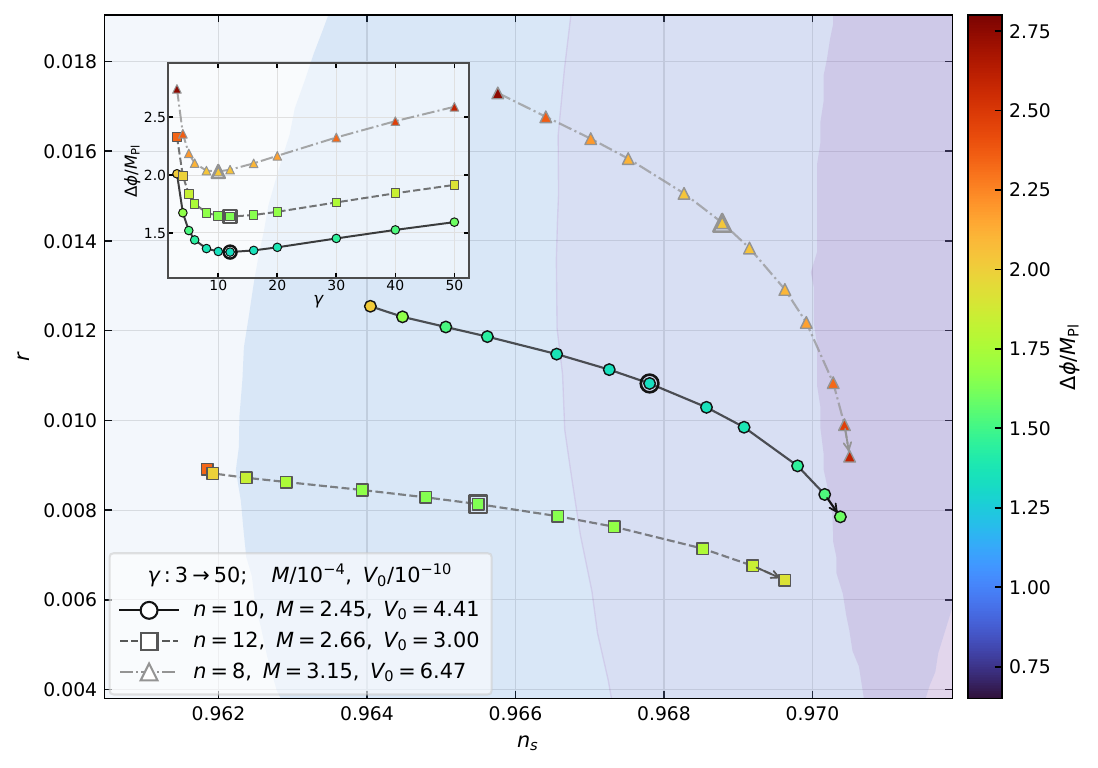}
    \caption{Left panel: effects of $M$ with the other parameters fixed. Right panel: effects of $\gamma$ with the other parameters fixed. Again, $\lambda$ is chosen so that $A_s = 2.1 \times 10^{-9}$ remains fixed. The data for the $n=10$ branch in the right panel are listed in Table~\ref{tab:B49} of Appendix \ref{app:model2}. The color bar indicates the original field excursion $\Delta \phi/M_p$.}
    \label{fig:para}
\end{figure}

We also find that varying $n$ and $V_0$ has only a limited effect on the results. Increasing $n$ leads to a slight change in $n_s$, whereas increasing $V_0$ primarily helps satisfy the Lyth-bound constraint, with only a limited effect on the ACT constraint. The value of $M$ has a negligible impact on the results, as shown in the left panel of Fig. \ref{fig:para}. The sensitivity to $N_*$ is more significant: increasing $N_*$ can move $n_s$ toward the ACT-preferred value, but this shift is accompanied by a larger accumulated field excursion. Thus, the conclusion is not driven by a single fiducial choice of $N_*$.

Given this negative result, it is natural to ask whether $\gamma$, the most important parameter in these models, can improve the robustness of the model. As discussed above, increasing $\gamma$ can both suppress $r$ and substantially change $\Delta \phi$, so one may expect a larger $\gamma$ to help evade the Lyth-bound and ACT constraints. Nonetheless, $\gamma$ cannot be arbitrarily large. The effective field theory of inflation suggests $c_s \gg 0.003$ to avoid a strong-coupling problem \cite{Cheung:2007st}, while observational constraints on primordial non-Gaussianities imply $c_s \geq 0.021$ (for the DBI inflation model, the constraint can be even more severe) \cite{Cabass:2022wjy}. In particular, the strong coupling scale of the Goldstone EFT behaves as
\begin{equation}
    \Lambda_{\rm sc}^{4}\simeq \frac{16 \pi^2 M_p^2|\dot H|c_s^5}{1-c_s^2}~,
\end{equation}
up to coefficients of order unity. Combining this estimate with $A_s=H^2/(8\pi^2M_p^2\epsilon c_s)$ yields
\begin{equation}
    \left(\frac{H}{\Lambda_{\rm sc}}\right)^4\simeq \frac{A_s}{2}\frac{1-c_s^2}{c_s^4}~,
\end{equation}
Thus, $c_s\gtrsim0.1$ keeps $H/\Lambda_{\rm sc}\ll1$, whereas reducing the sound speed substantially would make the EFT interpretation less robust and enhance equilateral non-Gaussianities. Therefore, we adopt $\gamma \lesssim 50$ to ensure $c_s \geq 0.1$.

In light of this discussion, we vary the parameter $\gamma$ while keeping $n$, $V_0$, and $N$ fixed, and present the corresponding values of $n_s, r$, and $\Delta \phi$ in the right panel of Fig. \ref{fig:para}. The data for the $n=10$ branch in this panel are listed in Table~\ref{tab:B49} of Appendix \ref{app:model2}. Interestingly, a lower bound on $\Delta \phi$ that exceeds $M_p$ emerges when $\gamma$ is varied, suggesting that changing $\gamma$ alone cannot circumvent the Lyth bound constraints.

The emergence of a lower bound can be explained as follows. To demonstrate this, we evaluate $d\Delta \phi/d \gamma$. The starting point is
\begin{equation}
\label{eq:dphidgamma}
    \frac{d\Delta \phi}{d \gamma} = \frac{d \phi_{\rm end}}{d \gamma} - \frac{d\phi_{\ast}}{d \gamma} ~.
\end{equation}
The right-hand side is determined by taking the total derivatives of the constraints.
\begin{equation}
\label{eq:impfun}
    P_{\zeta} (\phi_{\ast},\gamma, \lambda) = A_s ~,~ \epsilon (\phi_{\rm end} ,\gamma, \lambda) = 1 ~.
\end{equation}
Because $N_{\ast}$ is fixed, an additional constraint must be considered.
\begin{equation}
    \int_{\phi_{\ast}}^{\phi_{\mathrm{end}}} [\frac{d\phi}{dN} (\gamma,\lambda)]^{-1} d \phi=N_{\ast} ~.
\end{equation}
Substituting these expressions into Eq. \eqref{eq:dphidgamma}, we obtain
\begin{equation}
    \frac{d\Delta \phi}{d\gamma}= D_\gamma+ C_\lambda \frac{d\lambda}{d\gamma} ~,     D_\gamma = -\frac{\partial_\gamma \epsilon}{\partial_{\phi_{\rm end}} \epsilon}+\frac{\partial_\gamma P_s}{\partial_{\phi_{\ast}} P_s} ~,~ C_\lambda=-\frac{\partial_\lambda \epsilon}{\partial_{\phi_{\rm end}} \epsilon}+\frac{\partial_\lambda P_s}{\partial_{\phi_{\ast}} P_s} ~.
\end{equation}
Note that $P_s$ and $\epsilon$ are implicit functions defined by Eq. \eqref{eq:impfun}. Because a complete analytical treatment is intractable, we proceed with a numerical evaluation. We list the values of $d\lambda /d\gamma$, $D_{\gamma}$, and $C_{\lambda}$ in Table \ref{tab:gamma_derivative_decomposition}. For this parameter choice, $D_{\gamma}$ increases from negative values toward 0 as $\gamma$ increases. In contrast, $C_\lambda \frac{d\lambda}{d\gamma}$ eventually becomes positive as $\gamma$ increases. This behavior leads to a sign reversal in $d\Delta \phi/d\gamma$ and, consequently, to a local minimum in $\Delta \phi$ as $\gamma$ is varied. 

\begin{table}[htbp]
\centering
\begin{tabular}{ccccc}
\toprule
$\gamma$
& $d\lambda/d\gamma$
& $D_\gamma$ 
& $C_\lambda d\lambda/d\gamma$
& $\frac{d\Delta\phi}{d\gamma}$
\\
\midrule
1.5
& $+1.5802\times 10^{-10}$
& $-0.2733$
& $-5.8120$
& $-6.0853$
\\
6
& $+1.6722\times 10^{-7}$
& $-0.0324$
& $-0.0280$
& $-0.0605$
\\
8
& $-6.5174\times 10^{-9}$
& $-0.0220$
& $+0.00087$
& $-0.0211$
\\
12
& $-7.7776\times 10^{-8}$
& $-0.0135$
& $+0.01393$
& $+0.00042$
\\
50
& $-1.0290\times 10^{-9}$
& $-0.00332$
& $+0.00972$
& $+0.00640$
\\
\bottomrule
\end{tabular}
\caption{Evolution of $\Delta\phi$ along the $\gamma$ branch.}
\label{tab:gamma_derivative_decomposition}
\end{table}

In this regard, the model may have difficulty simultaneously evading the Lyth-bound constraint and fitting the ACT observations by adjusting its parameters. We therefore reiterate that the joint consideration of the Lyth bound and ACT results can effectively constrain inflationary scenarios, placing models that are compatible with either constraint alone in tension.

\section{Conclusion}
\label{sec:conclusion}

We have investigated how the most recent ACT observations shed new light on the physics of the Lyth bound. Although some inflationary models may remain viable under the original Lyth bound, the inclusion of a red tilt can strengthen the constraint and place such models in tension with observations. This is explicitly demonstrated by the kinetic-inflaton coupling model, which was designed to evade the Lyth-bound constraint: the Lyth bound motivates the construction of such a model, while the inclusion of the red tilt suggests that its predictions for $(n_s,r)$ are incompatible with the ACT observations. We also investigated models with a modified kinetic term and found that they can satisfy the joint Planck 2018 and Lyth-bound constraints, but fail when the Planck 2018 data are replaced by the ACT data.

Given the abundance of inflationary models in the literature, an exhaustive investigation of all existing scenarios is not possible. The purpose of this manuscript is to highlight the power of jointly applying the theoretical Lyth-bound constraint and the most up-to-date CMB observations. We have already found this method to be powerful, and extending the formalism to other models in future work will be promising.

\begin{acknowledgments}
We thank Yizhi Liang, Guohe Li and Jingming Pu for useful discussions and insightful suggestions. 
\end{acknowledgments}

\appendix
\section{Data for Figure 1}
\label{Model I Data}
\setcounter{table}{0}
\renewcommand{\thetable}{A\arabic{table}}

This section summarizes the data underlying Fig. \ref{fig:nsrI}. We assume $\alpha=30$ throughout. We present both the numerical results and the semi-analytical results obtained using the slow-roll approximation.

\begingroup
\tiny
\setlength{\tabcolsep}{2.2pt}
\renewcommand{\arraystretch}{1.02}

\noindent\begin{minipage}[t]{0.48\textwidth}
\appendixtablecaption{tab:A1}{Numerical, $N_*=50$.}
\begin{tabular}{@{}ccccc@{}}
\toprule
$\tilde{\alpha}$ & $\beta$ & $n_s$ & $r$ & $\Delta\phi/M_{\rm Pl}$ \\
\midrule
0.005 & 0.0019965 & 0.959698 & 0.158314 & 0.17575 \\
0.010167 & \ensuremath{9.8189\times10^{-4}} & 0.9599148 & 0.158631 & 0.175813 \\
0.015333 & \ensuremath{6.5103\times10^{-4}} & 0.9596259 & 0.159161 & 0.17592 \\
0.0205 & \ensuremath{4.8700\times10^{-4}} & 0.9598288 & 0.159903 & 0.176071 \\
0.025667 & \ensuremath{3.8898\times10^{-4}} & 0.9595938 & 0.160862 & 0.176265 \\
0.030833 & \ensuremath{3.2379\times10^{-4}} & 0.9595782 & 0.162039 & 0.176504 \\
0.036 & \ensuremath{2.7730\times10^{-4}} & 0.9596289 & 0.163435 & 0.176786 \\
\bottomrule
\end{tabular}
\end{minipage}\hfill
\begin{minipage}[t]{0.48\textwidth}
\appendixtablecaption{tab:A2}{Numerical, $N_*=60$.}
\begin{tabular}{@{}ccccc@{}}
\toprule
$\tilde{\alpha}$ & $\beta$ & $n_s$ & $r$ & $\Delta\phi/M_{\rm Pl}$ \\
\midrule
0.005 & 0.0016643 & 0.966458 & 0.132254 & 0.181873 \\
0.010167 & \ensuremath{8.1847\times10^{-4}} & 0.9664713 & 0.132573 & 0.18195 \\
0.015333 & \ensuremath{5.4274\times10^{-4}} & 0.9664267 & 0.133096 & 0.182079 \\
0.0205 & \ensuremath{4.0596\times10^{-4}} & 0.9666084 & 0.133848 & 0.18226 \\
0.025667 & \ensuremath{3.2426\times10^{-4}} & 0.9666834 & 0.134819 & 0.182494 \\
0.030833 & \ensuremath{2.6991\times10^{-4}} & 0.9665007 & 0.135997 & 0.182782 \\
0.036 & \ensuremath{2.3116\times10^{-4}} & 0.9662986 & 0.137408 & 0.183123 \\
\bottomrule
\end{tabular}
\end{minipage}
\par\medskip

\noindent\begin{minipage}[t]{0.48\textwidth}
\appendixtablecaption{tab:A3}{Numerical, $N_*=70$.}
\begin{tabular}{@{}ccccc@{}}
\toprule
$\tilde{\alpha}$ & $\beta$ & $n_s$ & $r$ & $\Delta\phi/M_{\rm Pl}$ \\
\midrule
0.005 & 0.0014269 & 0.9712366 & 0.113558 & 0.187047 \\
0.010167 & \ensuremath{7.0175\times10^{-4}} & 0.97132 & 0.11387 & 0.187137 \\
0.015333 & \ensuremath{4.6533\times10^{-4}} & 0.971561 & 0.114407 & 0.187288 \\
0.0205 & \ensuremath{3.4803\times10^{-4}} & 0.9711759 & 0.115136 & 0.187501 \\
0.025667 & \ensuremath{2.7798\times10^{-4}} & 0.97098 & 0.116122 & 0.187775 \\
0.030833 & \ensuremath{2.3139\times10^{-4}} & 0.971168 & 0.117301 & 0.188112 \\
0.036 & \ensuremath{1.9815\times10^{-4}} & 0.9711718 & 0.118704 & 0.188512 \\
\bottomrule
\end{tabular}
\end{minipage}\hfill
\begin{minipage}[t]{0.48\textwidth}
\appendixtablecaption{tab:A4}{Numerical, $N_*=80$.}
\begin{tabular}{@{}ccccc@{}}
\toprule
$\tilde{\alpha}$ & $\beta$ & $n_s$ & $r$ & $\Delta\phi/M_{\rm Pl}$ \\
\midrule
0.005 & 0.0012486 & 0.9749345 & 0.0994967 & 0.191528 \\
0.010167 & \ensuremath{6.1406\times10^{-4}} & 0.9748769 & 0.0997936 & 0.191631 \\
0.015333 & \ensuremath{4.0720\times10^{-4}} & 0.9751382 & 0.10032 & 0.191804 \\
0.0205 & \ensuremath{3.0452\times10^{-4}} & 0.9750585 & 0.101067 & 0.192047 \\
0.025667 & \ensuremath{2.4324\times10^{-4}} & 0.9748235 & 0.102031 & 0.192362 \\
0.030833 & \ensuremath{2.0246\times10^{-4}} & 0.9748628 & 0.10322 & 0.192749 \\
0.036 & \ensuremath{1.7338\times10^{-4}} & 0.9749764 & 0.104653 & 0.193208 \\
\bottomrule
\end{tabular}
\end{minipage}
\par\medskip

\noindent\begin{minipage}[t]{0.48\textwidth}
\appendixtablecaption{tab:A5}{Semi-analytic, $N_*=50$.}
\begin{tabular}{@{}ccccc@{}}
\toprule
$\tilde{\alpha}$ & $\beta$ & $n_s$ & $r$ & $\Delta\phi/M_{\rm Pl}$ \\
\midrule
0.005 & 0.0019753 & 0.9603924 & 0.158532 & 0.153928 \\
0.010167 & \ensuremath{9.7147\times10^{-4}} & 0.9603924 & 0.158846 & 0.153993 \\
0.015333 & \ensuremath{6.4412\times10^{-4}} & 0.960392 & 0.159374 & 0.154102 \\
0.0205 & \ensuremath{4.8178\times10^{-4}} & 0.960391 & 0.160119 & 0.154255 \\
0.025667 & \ensuremath{3.8478\times10^{-4}} & 0.9603889 & 0.161081 & 0.154452 \\
0.030833 & \ensuremath{3.2029\times10^{-4}} & 0.9603849 & 0.162265 & 0.154695 \\
0.036 & \ensuremath{2.7430\times10^{-4}} & 0.9603784 & 0.163672 & 0.154983 \\
\bottomrule
\end{tabular}
\end{minipage}\hfill
\begin{minipage}[t]{0.48\textwidth}
\appendixtablecaption{tab:A6}{Semi-analytic, $N_*=60$.}
\begin{tabular}{@{}ccccc@{}}
\toprule
$\tilde{\alpha}$ & $\beta$ & $n_s$ & $r$ & $\Delta\phi/M_{\rm Pl}$ \\
\midrule
0.005 & 0.0016488 & 0.9669395 & 0.132343 & 0.159955 \\
0.010167 & \ensuremath{8.1088\times10^{-4}} & 0.9669394 & 0.132657 & 0.160033 \\
0.015333 & \ensuremath{5.3765\times10^{-4}} & 0.966939 & 0.133186 & 0.160164 \\
0.0205 & \ensuremath{4.0213\times10^{-4}} & 0.9669378 & 0.133931 & 0.160348 \\
0.025667 & \ensuremath{3.2117\times10^{-4}} & 0.9669352 & 0.134895 & 0.160586 \\
0.030833 & \ensuremath{2.6733\times10^{-4}} & 0.9669305 & 0.136082 & 0.160878 \\
0.036 & \ensuremath{2.2894\times10^{-4}} & 0.9669227 & 0.137494 & 0.161224 \\
\bottomrule
\end{tabular}
\end{minipage}
\par\medskip

\noindent\begin{minipage}[t]{0.48\textwidth}
\appendixtablecaption{tab:A7}{Semi-analytic, $N_*=70$.}
\begin{tabular}{@{}ccccc@{}}
\toprule
$\tilde{\alpha}$ & $\beta$ & $n_s$ & $r$ & $\Delta\phi/M_{\rm Pl}$ \\
\midrule
0.005 & 0.0014149 & 0.9716292 & 0.113584 & 0.165059 \\
0.010167 & \ensuremath{6.9585\times10^{-4}} & 0.9716291 & 0.113898 & 0.165149 \\
0.015333 & \ensuremath{4.6138\times10^{-4}} & 0.9716286 & 0.114427 & 0.165302 \\
0.0205 & \ensuremath{3.4509\times10^{-4}} & 0.9716272 & 0.115173 & 0.165518 \\
0.025667 & \ensuremath{2.7561\times10^{-4}} & 0.9716241 & 0.116139 & 0.165796 \\
0.030833 & \ensuremath{2.2940\times10^{-4}} & 0.9716186 & 0.117329 & 0.166137 \\
0.036 & \ensuremath{1.9645\times10^{-4}} & 0.9716095 & 0.118747 & 0.166542 \\
\bottomrule
\end{tabular}
\end{minipage}\hfill
\begin{minipage}[t]{0.48\textwidth}
\appendixtablecaption{tab:A8}{Semi-analytic, $N_*=80$.}
\begin{tabular}{@{}ccccc@{}}
\toprule
$\tilde{\alpha}$ & $\beta$ & $n_s$ & $r$ & $\Delta\phi/M_{\rm Pl}$ \\
\midrule
0.005 & 0.0012391 & 0.9751537 & 0.0994859 & 0.169485 \\
0.010167 & \ensuremath{6.0941\times10^{-4}} & 0.9751536 & 0.0997999 & 0.169589 \\
0.015333 & \ensuremath{4.0406\times10^{-4}} & 0.975153 & 0.100329 & 0.169764 \\
0.0205 & \ensuremath{3.0221\times10^{-4}} & 0.9751514 & 0.101076 & 0.17001 \\
0.025667 & \ensuremath{2.4136\times10^{-4}} & 0.9751479 & 0.102044 & 0.170328 \\
0.030833 & \ensuremath{2.0089\times10^{-4}} & 0.9751416 & 0.103237 & 0.170719 \\
0.036 & \ensuremath{1.7202\times10^{-4}} & 0.9751312 & 0.10466 & 0.171184 \\
\bottomrule
\end{tabular}
\end{minipage}
\par\medskip
\endgroup

\section{Data for the modified kinetic term model}

\label{app:model2}
\setcounter{table}{0}
\renewcommand{\thetable}{B\arabic{table}}
We list the points used to plot Fig. \ref{fig:model2res} for the modified kinetic-term model, with $M \simeq 2.4\times10^{-4}$ fixed. We present both the numerical results and the semi-analytical results obtained using the slow-roll approximation. The first column gives $V_0/10^{-10}$.

\begingroup
\tiny
\setlength{\tabcolsep}{2pt}
\renewcommand{\arraystretch}{1.02}

\subsection*{Numerical points}
\label{tab:model2-numerical-figure-points-compact}

\noindent\begin{minipage}[t]{0.48\textwidth}
\appendixtablecaption{tab:B1}{$\gamma=10$, $n=8$, $N=50$.}
\begin{tabular}{@{}cccccc@{}}
\toprule
$V_0$ & $\lambda$ & $n_s$ & $r$ & $\Delta\phi/M_{\rm Pl}$ & $H/M$ \\
\midrule
3 & 0.021058 & 0.931357 & 0.00885 & 0.86008 & 0.039266 \\
4.4054 & 0.022844 & 0.939657 & 0.012296 & 0.78524 & 0.046366 \\
6.4693 & 0.019655 & 0.945116 & 0.016545 & 0.7318 & 0.053901 \\
9.5 & 0.013541 & 0.948814 & 0.021338 & 0.69741 & 0.061366 \\
\bottomrule
\end{tabular}
\end{minipage}\hfill
\begin{minipage}[t]{0.48\textwidth}
\appendixtablecaption{tab:B2}{$\gamma=10$, $n=8$, $N=60$.}
\begin{tabular}{@{}cccccc@{}}
\toprule
$V_0$ & $\lambda$ & $n_s$ & $r$ & $\Delta\phi/M_{\rm Pl}$ & $H/M$ \\
\midrule
3 & 0.0029811 & 0.946125 & 0.0086741 & 1.0745 & 0.038866 \\
4.4054 & 0.0028968 & 0.951631 & 0.011898 & 0.99107 & 0.045594 \\
6.4693 & 0.0022291 & 0.955733 & 0.015724 & 0.93438 & 0.052517 \\
9.5 & 0.0013993 & 0.958068 & 0.019852 & 0.90065 & 0.059134 \\
\bottomrule
\end{tabular}
\end{minipage}
\par\medskip

\noindent\begin{minipage}[t]{0.48\textwidth}
\appendixtablecaption{tab:B3}{$\gamma=10$, $n=8$, $N=70$.}
\begin{tabular}{@{}cccccc@{}}
\toprule
$V_0$ & $\lambda$ & $n_s$ & $r$ & $\Delta\phi/M_{\rm Pl}$ & $H/M$ \\
\midrule
3 & 5.4643e-04 & 0.955908 & 0.0084912 & 1.3017 & 0.038447 \\
4.4054 & 4.8253e-04 & 0.96018 & 0.011498 & 1.2118 & 0.044807 \\
6.4693 & 3.3948e-04 & 0.963017 & 0.014941 & 1.1539 & 0.051166 \\
9.5 & 1.9931e-04 & 0.964599 & 0.018516 & 1.1223 & 0.057064 \\
\bottomrule
\end{tabular}
\end{minipage}\hfill
\begin{minipage}[t]{0.48\textwidth}
\appendixtablecaption{tab:B4}{$\gamma=10$, $n=8$, $N=80$.}
\begin{tabular}{@{}cccccc@{}}
\toprule
$V_0$ & $\lambda$ & $n_s$ & $r$ & $\Delta\phi/M_{\rm Pl}$ & $H/M$ \\
\midrule
3 & 1.2157e-04 & 0.962842 & 0.008304 & 1.5414 & 0.038016 \\
4.4054 & 9.8823e-05 & 0.966155 & 0.011103 & 1.4471 & 0.044019 \\
6.4693 & 6.4710e-05 & 0.968161 & 0.014204 & 1.3897 & 0.049867 \\
9.5 & 3.6206e-05 & 0.96934 & 0.017323 & 1.3613 & 0.055157 \\
\bottomrule
\end{tabular}
\end{minipage}
\par\medskip

\noindent\begin{minipage}[t]{0.48\textwidth}
\appendixtablecaption{tab:B5}{$\gamma=10$, $n=10$, $N=50$.}
\begin{tabular}{@{}cccccc@{}}
\toprule
$V_0$ & $\lambda$ & $n_s$ & $r$ & $\Delta\phi/M_{\rm Pl}$ & $H/M$ \\
\midrule
3 & 8.8905e-04 & 0.934014 & 0.0088094 & 0.84912 & 0.039174 \\
4.4054 & 9.1425e-04 & 0.941325 & 0.012216 & 0.77667 & 0.046213 \\
6.4693 & 7.0744e-04 & 0.94595 & 0.016402 & 0.72503 & 0.053664 \\
9.5 & 4.1845e-04 & 0.949302 & 0.021113 & 0.69187 & 0.061033 \\
\bottomrule
\end{tabular}
\end{minipage}\hfill
\begin{minipage}[t]{0.48\textwidth}
\appendixtablecaption{tab:B6}{$\gamma=10$, $n=10$, $N=60$.}
\begin{tabular}{@{}cccccc@{}}
\toprule
$V_0$ & $\lambda$ & $n_s$ & $r$ & $\Delta\phi/M_{\rm Pl}$ & $H/M$ \\
\midrule
3 & 7.4569e-05 & 0.947793 & 0.0086269 & 1.0613 & 0.038758 \\
4.4054 & 6.6943e-05 & 0.952748 & 0.011808 & 0.98063 & 0.045419 \\
6.4693 & 4.5223e-05 & 0.956414 & 0.015572 & 0.92599 & 0.052258 \\
9.5 & 2.3939e-05 & 0.958596 & 0.019627 & 0.89353 & 0.058791 \\
\bottomrule
\end{tabular}
\end{minipage}
\par\medskip

\noindent\begin{minipage}[t]{0.48\textwidth}
\appendixtablecaption{tab:B7}{$\gamma=10$, $n=10$, $N=70$.}
\begin{tabular}{@{}cccccc@{}}
\toprule
$V_0$ & $\lambda$ & $n_s$ & $r$ & $\Delta\phi/M_{\rm Pl}$ & $H/M$ \\
\midrule
3 & 8.6873e-06 & 0.957014 & 0.008438 & 1.2861 & 0.038325 \\
4.4054 & 6.9357e-06 & 0.960904 & 0.011401 & 1.1994 & 0.044616 \\
6.4693 & 4.2054e-06 & 0.963614 & 0.014785 & 1.1437 & 0.050895 \\
9.5 & 2.0568e-06 & 0.964989 & 0.018297 & 1.1134 & 0.056719 \\
\bottomrule
\end{tabular}
\end{minipage}\hfill
\begin{minipage}[t]{0.48\textwidth}
\appendixtablecaption{tab:B8}{$\gamma=10$, $n=10$, $N=80$.}
\begin{tabular}{@{}cccccc@{}}
\toprule
$V_0$ & $\lambda$ & $n_s$ & $r$ & $\Delta\phi/M_{\rm Pl}$ & $H/M$ \\
\midrule
3 & 1.2949e-06 & 0.963727 & 0.0082458 & 1.5235 & 0.037881 \\
4.4054 & 9.3463e-07 & 0.966678 & 0.011001 & 1.4327 & 0.043814 \\
6.4693 & 5.2008e-07 & 0.968495 & 0.014048 & 1.3775 & 0.049587 \\
9.5 & 2.4039e-07 & 0.969618 & 0.017113 & 1.3504 & 0.054816 \\
\bottomrule
\end{tabular}
\end{minipage}
\par\medskip

\noindent\begin{minipage}[t]{0.48\textwidth}
\appendixtablecaption{tab:B9}{$\gamma=10$, $n=12$, $N=50$.}
\begin{tabular}{@{}cccccc@{}}
\toprule
$V_0$ & $\lambda$ & $n_s$ & $r$ & $\Delta\phi/M_{\rm Pl}$ & $H/M$ \\
\midrule
3 & 2.5050e-05 & 0.934897 & 0.0087823 & 0.8422 & 0.039113 \\
4.4054 & 2.4429e-05 & 0.941484 & 0.012164 & 0.77123 & 0.046111 \\
6.4693 & 1.7004e-05 & 0.946355 & 0.016308 & 0.72073 & 0.053507 \\
9.5 & 8.6395e-06 & 0.949571 & 0.020965 & 0.68833 & 0.060815 \\
\bottomrule
\end{tabular}
\end{minipage}\hfill
\begin{minipage}[t]{0.48\textwidth}
\appendixtablecaption{tab:B10}{$\gamma=10$, $n=12$, $N=60$.}
\begin{tabular}{@{}cccccc@{}}
\toprule
$V_0$ & $\lambda$ & $n_s$ & $r$ & $\Delta\phi/M_{\rm Pl}$ & $H/M$ \\
\midrule
3 & 1.2448e-06 & 0.948747 & 0.0085953 & 1.0529 & 0.038687 \\
4.4054 & 1.0328e-06 & 0.953546 & 0.011749 & 0.974 & 0.045304 \\
6.4693 & 6.1281e-07 & 0.956873 & 0.015473 & 0.92062 & 0.052088 \\
9.5 & 2.7370e-07 & 0.958941 & 0.019481 & 0.88895 & 0.058566 \\
\bottomrule
\end{tabular}
\end{minipage}
\par\medskip

\noindent\begin{minipage}[t]{0.48\textwidth}
\appendixtablecaption{tab:B11}{$\gamma=10$, $n=12$, $N=70$.}
\begin{tabular}{@{}cccccc@{}}
\toprule
$V_0$ & $\lambda$ & $n_s$ & $r$ & $\Delta\phi/M_{\rm Pl}$ & $H/M$ \\
\midrule
3 & 9.2173e-08 & 0.957822 & 0.0084027 & 1.2763 & 0.038244 \\
4.4054 & 6.6566e-08 & 0.961552 & 0.011337 & 1.1916 & 0.044489 \\
6.4693 & 3.4807e-08 & 0.963913 & 0.014683 & 1.1372 & 0.050716 \\
9.5 & 1.4186e-08 & 0.965165 & 0.018155 & 1.1076 & 0.056494 \\
\bottomrule
\end{tabular}
\end{minipage}\hfill
\begin{minipage}[t]{0.48\textwidth}
\appendixtablecaption{tab:B12}{$\gamma=10$, $n=12$, $N=80$.}
\begin{tabular}{@{}cccccc@{}}
\toprule
$V_0$ & $\lambda$ & $n_s$ & $r$ & $\Delta\phi/M_{\rm Pl}$ & $H/M$ \\
\midrule
3 & 9.2072e-09 & 0.9643 & 0.0082073 & 1.5121 & 0.037791 \\
4.4054 & 5.9035e-09 & 0.967135 & 0.010934 & 1.4235 & 0.043679 \\
6.4693 & 2.7932e-09 & 0.968944 & 0.013945 & 1.3697 & 0.049404 \\
9.5 & 1.0669e-09 & 0.969715 & 0.016976 & 1.3434 & 0.054592 \\
\bottomrule
\end{tabular}
\end{minipage}
\par\medskip

\noindent\begin{minipage}[t]{0.48\textwidth}
\appendixtablecaption{tab:B13}{$\gamma=12$, $n=8$, $N=50$.}
\begin{tabular}{@{}cccccc@{}}
\toprule
$V_0$ & $\lambda$ & $n_s$ & $r$ & $\Delta\phi/M_{\rm Pl}$ & $H/M$ \\
\midrule
4.4054 & 0.022031 & 0.941187 & 0.012081 & 0.77398 & 0.046002 \\
6.4693 & 0.018215 & 0.946138 & 0.016098 & 0.72329 & 0.053229 \\
9.5 & 0.012089 & 0.949198 & 0.020514 & 0.69164 & 0.060249 \\
\bottomrule
\end{tabular}
\end{minipage}\hfill
\begin{minipage}[t]{0.48\textwidth}
\appendixtablecaption{tab:B14}{$\gamma=12$, $n=8$, $N=60$.}
\begin{tabular}{@{}cccccc@{}}
\toprule
$V_0$ & $\lambda$ & $n_s$ & $r$ & $\Delta\phi/M_{\rm Pl}$ & $H/M$ \\
\midrule
4.4054 & 0.0026752 & 0.952769 & 0.011645 & 0.98111 & 0.045145 \\
6.4693 & 0.001977 & 0.95638 & 0.015222 & 0.92797 & 0.051725 \\
9.5 & 0.0012012 & 0.958461 & 0.018981 & 0.89748 & 0.057889 \\
\bottomrule
\end{tabular}
\end{minipage}
\par\medskip

\noindent\begin{minipage}[t]{0.48\textwidth}
\appendixtablecaption{tab:B15}{$\gamma=12$, $n=8$, $N=70$.}
\begin{tabular}{@{}cccccc@{}}
\toprule
$V_0$ & $\lambda$ & $n_s$ & $r$ & $\Delta\phi/M_{\rm Pl}$ & $H/M$ \\
\midrule
4.4054 & 4.2921e-04 & 0.960895 & 0.011211 & 1.2043 & 0.04428 \\
6.4693 & 2.9049e-04 & 0.963483 & 0.014398 & 1.1507 & 0.050275 \\
9.5 & 1.6597e-04 & 0.964898 & 0.017622 & 1.1228 & 0.055726 \\
\bottomrule
\end{tabular}
\end{minipage}\hfill
\begin{minipage}[t]{0.48\textwidth}
\appendixtablecaption{tab:B16}{$\gamma=12$, $n=8$, $N=80$.}
\begin{tabular}{@{}cccccc@{}}
\toprule
$V_0$ & $\lambda$ & $n_s$ & $r$ & $\Delta\phi/M_{\rm Pl}$ & $H/M$ \\
\midrule
4.4054 & 8.5089e-05 & 0.966759 & 0.010787 & 1.4432 & 0.04342 \\
6.4693 & 5.3764e-05 & 0.968638 & 0.013632 & 1.3907 & 0.048893 \\
9.5 & 2.9423e-05 & 0.969549 & 0.016424 & 1.3664 & 0.053755 \\
\bottomrule
\end{tabular}
\end{minipage}
\par\medskip

\noindent\begin{minipage}[t]{0.48\textwidth}
\appendixtablecaption{tab:B17}{$\gamma=12$, $n=10$, $N=50$.}
\begin{tabular}{@{}cccccc@{}}
\toprule
$V_0$ & $\lambda$ & $n_s$ & $r$ & $\Delta\phi/M_{\rm Pl}$ & $H/M$ \\
\midrule
4.4054 & 8.5905e-04 & 0.941387 & 0.011997 & 0.76592 & 0.045838 \\
6.4693 & 6.3342e-04 & 0.946882 & 0.015951 & 0.71693 & 0.052981 \\
9.5 & 3.5844e-04 & 0.949956 & 0.020291 & 0.68639 & 0.059912 \\
\bottomrule
\end{tabular}
\end{minipage}\hfill
\begin{minipage}[t]{0.48\textwidth}
\appendixtablecaption{tab:B18}{$\gamma=12$, $n=10$, $N=60$.}
\begin{tabular}{@{}cccccc@{}}
\toprule
$V_0$ & $\lambda$ & $n_s$ & $r$ & $\Delta\phi/M_{\rm Pl}$ & $H/M$ \\
\midrule
4.4054 & 5.9614e-05 & 0.953875 & 0.011552 & 0.97127 & 0.044961 \\
6.4693 & 3.8370e-05 & 0.95705 & 0.015069 & 0.92001 & 0.051459 \\
9.5 & 1.9553e-05 & 0.959008 & 0.018762 & 0.89064 & 0.057546 \\
\bottomrule
\end{tabular}
\end{minipage}
\par\medskip

\noindent\begin{minipage}[t]{0.48\textwidth}
\appendixtablecaption{tab:B19}{$\gamma=12$, $n=10$, $N=70$.}
\begin{tabular}{@{}cccccc@{}}
\toprule
$V_0$ & $\lambda$ & $n_s$ & $r$ & $\Delta\phi/M_{\rm Pl}$ & $H/M$ \\
\midrule
4.4054 & 5.8977e-06 & 0.961737 & 0.011111 & 1.1926 & 0.044079 \\
6.4693 & 3.4157e-06 & 0.964046 & 0.014243 & 1.141 & 0.049998 \\
9.5 & 1.6188e-06 & 0.965173 & 0.017412 & 1.1141 & 0.055385 \\
\bottomrule
\end{tabular}
\end{minipage}\hfill
\begin{minipage}[t]{0.48\textwidth}
\appendixtablecaption{tab:B20}{$\gamma=12$, $n=10$, $N=80$.}
\begin{tabular}{@{}cccccc@{}}
\toprule
$V_0$ & $\lambda$ & $n_s$ & $r$ & $\Delta\phi/M_{\rm Pl}$ & $H/M$ \\
\midrule
4.4054 & 7.6373e-07 & 0.96733 & 0.010683 & 1.4295 & 0.043207 \\
6.4693 & 4.0751e-07 & 0.968998 & 0.013478 & 1.379 & 0.048611 \\
9.5 & 1.8366e-07 & 0.969752 & 0.016223 & 1.3557 & 0.053419 \\
\bottomrule
\end{tabular}
\end{minipage}
\par\medskip

\noindent\begin{minipage}[t]{0.48\textwidth}
\appendixtablecaption{tab:B21}{$\gamma=12$, $n=12$, $N=50$.}
\begin{tabular}{@{}cccccc@{}}
\toprule
$V_0$ & $\lambda$ & $n_s$ & $r$ & $\Delta\phi/M_{\rm Pl}$ & $H/M$ \\
\midrule
4.4054 & 2.2365e-05 & 0.943329 & 0.011942 & 0.76081 & 0.04573 \\
6.4693 & 1.4711e-05 & 0.947593 & 0.015855 & 0.71288 & 0.052818 \\
9.5 & 7.1038e-06 & 0.950279 & 0.020146 & 0.68301 & 0.059691 \\
\bottomrule
\end{tabular}
\end{minipage}\hfill
\begin{minipage}[t]{0.48\textwidth}
\appendixtablecaption{tab:B22}{$\gamma=12$, $n=12$, $N=60$.}
\begin{tabular}{@{}cccccc@{}}
\toprule
$V_0$ & $\lambda$ & $n_s$ & $r$ & $\Delta\phi/M_{\rm Pl}$ & $H/M$ \\
\midrule
4.4054 & 8.8700e-07 & 0.954471 & 0.01149 & 0.96501 & 0.044839 \\
6.4693 & 4.9751e-07 & 0.957372 & 0.014969 & 0.91492 & 0.051284 \\
9.5 & 2.1271e-07 & 0.959237 & 0.018619 & 0.88624 & 0.057321 \\
\bottomrule
\end{tabular}
\end{minipage}
\par\medskip

\noindent\begin{minipage}[t]{0.48\textwidth}
\appendixtablecaption{tab:B23}{$\gamma=12$, $n=12$, $N=70$.}
\begin{tabular}{@{}cccccc@{}}
\toprule
$V_0$ & $\lambda$ & $n_s$ & $r$ & $\Delta\phi/M_{\rm Pl}$ & $H/M$ \\
\midrule
4.4054 & 5.4120e-08 & 0.962185 & 0.011046 & 1.1851 & 0.043947 \\
6.4693 & 2.6836e-08 & 0.964363 & 0.014141 & 1.1347 & 0.049816 \\
9.5 & 1.0553e-08 & 0.965515 & 0.017275 & 1.1085 & 0.055162 \\
\bottomrule
\end{tabular}
\end{minipage}\hfill
\begin{minipage}[t]{0.48\textwidth}
\appendixtablecaption{tab:B24}{$\gamma=12$, $n=12$, $N=80$.}
\begin{tabular}{@{}cccccc@{}}
\toprule
$V_0$ & $\lambda$ & $n_s$ & $r$ & $\Delta\phi/M_{\rm Pl}$ & $H/M$ \\
\midrule
4.4054 & 4.5786e-09 & 0.967677 & 0.010615 & 1.4207 & 0.043067 \\
6.4693 & 2.0642e-09 & 0.969228 & 0.013377 & 1.3715 & 0.048427 \\
9.5 & 7.6631e-10 & 0.96994 & 0.016093 & 1.3488 & 0.0532 \\
\bottomrule
\end{tabular}
\end{minipage}
\par\medskip

\subsection*{Semi-analytic Points}
\label{tab:model2-semi-figure-points-compact}

\noindent\begin{minipage}[t]{0.48\textwidth}
\appendixtablecaption{tab:B25}{$\gamma=10$, $n=8$, $N=50$.}
\begin{tabular}{@{}cccccc@{}}
\toprule
$V_0$ & $\lambda$ & $n_s$ & $r$ & $\Delta\phi/M_{\rm Pl}$ & $H/M$ \\
\midrule
3 & 0.01925 & 0.932736 & 0.0089542 & 0.80419 & 0.039318 \\
4.4054 & 0.020452 & 0.940342 & 0.012489 & 0.73196 & 0.046435 \\
6.4693 & 0.017132 & 0.946055 & 0.016874 & 0.68002 & 0.053974 \\
9.5 & 0.011444 & 0.949799 & 0.021845 & 0.6461 & 0.061412 \\
13.951 & 0.0064393 & 0.951838 & 0.026949 & 0.62671 & 0.06821 \\
20.486 & 0.0032937 & 0.952659 & 0.03176 & 0.61766 & 0.074049 \\
30.083 & 0.0016348 & 0.952744 & 0.036043 & 0.6152 & 0.078884 \\
44.177 & 8.2016e-04 & 0.952439 & 0.03974 & 0.61659 & 0.082831 \\
64.873 & 4.2440e-04 & 0.951949 & 0.042896 & 0.62012 & 0.086057 \\
95.264 & 2.2824e-04 & 0.951388 & 0.045587 & 0.62477 & 0.088716 \\
\bottomrule
\end{tabular}
\end{minipage}\hfill
\begin{minipage}[t]{0.48\textwidth}
\appendixtablecaption{tab:B26}{$\gamma=10$, $n=8$, $N=60$.}
\begin{tabular}{@{}cccccc@{}}
\toprule
$V_0$ & $\lambda$ & $n_s$ & $r$ & $\Delta\phi/M_{\rm Pl}$ & $H/M$ \\
\midrule
1.3912 & 0.0015532 & 0.931666 & 0.0043408 & 1.2526 & 0.027332 \\
2.0429 & 0.0022738 & 0.939867 & 0.0062101 & 1.1137 & 0.032744 \\
3 & 0.0027195 & 0.946964 & 0.0087692 & 1.0051 & 0.03891 \\
4.4054 & 0.0025896 & 0.952599 & 0.012068 & 0.92416 & 0.045645 \\
6.4693 & 0.001944 & 0.956591 & 0.016001 & 0.86856 & 0.052559 \\
9.5 & 0.0011884 & 0.959007 & 0.020257 & 0.83461 & 0.059137 \\
13.951 & 6.3388e-04 & 0.960179 & 0.024444 & 0.81713 & 0.064962 \\
20.486 & 3.1773e-04 & 0.960532 & 0.028272 & 0.81072 & 0.069864 \\
30.083 & 1.5800e-04 & 0.960421 & 0.031622 & 0.81103 & 0.073887 \\
44.177 & 8.0330e-05 & 0.960076 & 0.034493 & 0.81517 & 0.07717 \\
64.873 & 4.2321e-05 & 0.95963 & 0.036942 & 0.82134 & 0.079861 \\
95.264 & 2.3196e-05 & 0.959151 & 0.039034 & 0.8285 & 0.082092 \\
\bottomrule
\end{tabular}
\end{minipage}
\par\medskip

\noindent\begin{minipage}[t]{0.48\textwidth}
\appendixtablecaption{tab:B27}{$\gamma=10$, $n=8$, $N=70$.}
\begin{tabular}{@{}cccccc@{}}
\toprule
$V_0$ & $\lambda$ & $n_s$ & $r$ & $\Delta\phi/M_{\rm Pl}$ & $H/M$ \\
\midrule
0.94737 & 2.0968e-04 & 0.936586 & 0.0029736 & 1.6935 & 0.022658 \\
1.3912 & 3.3630e-04 & 0.944185 & 0.0042944 & 1.4969 & 0.027229 \\
2.0429 & 4.5449e-04 & 0.950917 & 0.0061291 & 1.3393 & 0.032529 \\
3 & 4.9834e-04 & 0.956539 & 0.0085781 & 1.218 & 0.038483 \\
4.4054 & 4.3160e-04 & 0.960826 & 0.011648 & 1.1303 & 0.044844 \\
6.4693 & 2.9685e-04 & 0.963695 & 0.015175 & 1.0729 & 0.051185 \\
9.5 & 1.7039e-04 & 0.965302 & 0.018845 & 1.0402 & 0.05704 \\
13.951 & 8.7993e-05 & 0.965986 & 0.02234 & 1.0254 & 0.062105 \\
20.486 & 4.3761e-05 & 0.9661 & 0.025471 & 1.0221 & 0.066313 \\
30.083 & 2.1906e-05 & 0.965906 & 0.028182 & 1.0256 & 0.069753 \\
44.177 & 1.1288e-05 & 0.965565 & 0.030498 & 1.0328 & 0.072563 \\
64.873 & 6.0414e-06 & 0.965165 & 0.032473 & 1.0419 & 0.074876 \\
95.264 & 3.3643e-06 & 0.964752 & 0.034165 & 1.0519 & 0.076801 \\
\bottomrule
\end{tabular}
\end{minipage}\hfill
\begin{minipage}[t]{0.48\textwidth}
\appendixtablecaption{tab:B28}{$\gamma=10$, $n=8$, $N=80$.}
\begin{tabular}{@{}cccccc@{}}
\toprule
$V_0$ & $\lambda$ & $n_s$ & $r$ & $\Delta\phi/M_{\rm Pl}$ & $H/M$ \\
\midrule
0.94737 & 5.7087e-05 & 0.946803 & 0.0029595 & 1.9718 & 0.022604 \\
1.3912 & 8.6703e-05 & 0.95322 & 0.0042596 & 1.7503 & 0.027118 \\
2.0429 & 1.0951e-04 & 0.958808 & 0.0060449 & 1.5749 & 0.032305 \\
3 & 1.1095e-04 & 0.96335 & 0.0083837 & 1.4426 & 0.038045 \\
4.4054 & 8.8532e-05 & 0.966677 & 0.011236 & 1.35 & 0.044044 \\
6.4693 & 5.6805e-05 & 0.968784 & 0.014404 & 1.2923 & 0.049868 \\
9.5 & 3.1177e-05 & 0.969876 & 0.017595 & 1.2619 & 0.055115 \\
13.951 & 1.5793e-05 & 0.970271 & 0.02056 & 1.2504 & 0.059579 \\
20.486 & 7.8444e-06 & 0.970257 & 0.023178 & 1.2507 & 0.063258 \\
30.083 & 3.9602e-06 & 0.97003 & 0.025431 & 1.2576 & 0.066261 \\
44.177 & 2.0668e-06 & 0.969707 & 0.027352 & 1.2682 & 0.068719 \\
64.873 & 1.1216e-06 & 0.969348 & 0.028993 & 1.2805 & 0.07075 \\
95.264 & 6.3307e-07 & 0.968987 & 0.030401 & 1.2935 & 0.072447 \\
\bottomrule
\end{tabular}
\end{minipage}
\par\medskip

\noindent\begin{minipage}[t]{0.48\textwidth}
\appendixtablecaption{tab:B29}{$\gamma=10$, $n=10$, $N=50$.}
\begin{tabular}{@{}cccccc@{}}
\toprule
$V_0$ & $\lambda$ & $n_s$ & $r$ & $\Delta\phi/M_{\rm Pl}$ & $H/M$ \\
\midrule
3 & 7.9674e-04 & 0.934977 & 0.0089122 & 0.79573 & 0.039226 \\
4.4054 & 7.9793e-04 & 0.941878 & 0.012406 & 0.72549 & 0.04628 \\
6.4693 & 5.9685e-04 & 0.947069 & 0.016724 & 0.67507 & 0.053734 \\
9.5 & 3.3975e-04 & 0.950468 & 0.021607 & 0.64215 & 0.061077 \\
13.951 & 1.5785e-04 & 0.952304 & 0.026622 & 0.62331 & 0.067795 \\
20.486 & 6.5702e-05 & 0.953019 & 0.03136 & 0.61451 & 0.073581 \\
30.083 & 2.6505e-05 & 0.953051 & 0.035593 & 0.61212 & 0.07839 \\
44.177 & 1.0882e-05 & 0.952717 & 0.039263 & 0.61352 & 0.082332 \\
64.873 & 4.6575e-06 & 0.95221 & 0.042408 & 0.61703 & 0.085566 \\
95.264 & 2.0963e-06 & 0.951638 & 0.045099 & 0.62167 & 0.08824 \\
\bottomrule
\end{tabular}
\end{minipage}\hfill
\begin{minipage}[t]{0.48\textwidth}
\appendixtablecaption{tab:B30}{$\gamma=10$, $n=10$, $N=60$.}
\begin{tabular}{@{}cccccc@{}}
\toprule
$V_0$ & $\lambda$ & $n_s$ & $r$ & $\Delta\phi/M_{\rm Pl}$ & $H/M$ \\
\midrule
1.3912 & 3.8760e-05 & 0.934667 & 0.0043165 & 1.2351 & 0.027299 \\
2.0429 & 5.7397e-05 & 0.942118 & 0.0061864 & 1.1004 & 0.032681 \\
3 & 6.6599e-05 & 0.948543 & 0.0087204 & 0.99484 & 0.038801 \\
4.4054 & 5.8276e-05 & 0.953661 & 0.011975 & 0.91628 & 0.045469 \\
6.4693 & 3.8168e-05 & 0.957287 & 0.015843 & 0.86238 & 0.052299 \\
9.5 & 1.9553e-05 & 0.959476 & 0.020022 & 0.82948 & 0.058794 \\
13.951 & 8.5371e-06 & 0.960521 & 0.024139 & 0.81252 & 0.064556 \\
20.486 & 3.4762e-06 & 0.960809 & 0.027915 & 0.8063 & 0.069422 \\
30.083 & 1.4087e-06 & 0.960665 & 0.031233 & 0.80667 & 0.073432 \\
44.177 & 5.8895e-07 & 0.960302 & 0.034089 & 0.8108 & 0.076716 \\
64.873 & 2.5807e-07 & 0.959844 & 0.036533 & 0.81695 & 0.079418 \\
95.264 & 1.1905e-07 & 0.959356 & 0.038627 & 0.8241 & 0.081663 \\
\bottomrule
\end{tabular}
\end{minipage}
\par\medskip

\noindent\begin{minipage}[t]{0.48\textwidth}
\appendixtablecaption{tab:B31}{$\gamma=10$, $n=10$, $N=70$.}
\begin{tabular}{@{}cccccc@{}}
\toprule
$V_0$ & $\lambda$ & $n_s$ & $r$ & $\Delta\phi/M_{\rm Pl}$ & $H/M$ \\
\midrule
0.94737 & 3.2803e-06 & 0.939715 & 0.0029677 & 1.6683 & 0.022635 \\
1.3912 & 5.5055e-06 & 0.946529 & 0.0042815 & 1.4769 & 0.027188 \\
2.0429 & 7.4438e-06 & 0.952606 & 0.0061017 & 1.3237 & 0.032457 \\
3 & 7.7525e-06 & 0.957704 & 0.0085234 & 1.2059 & 0.038361 \\
4.4054 & 6.0397e-06 & 0.961598 & 0.011548 & 1.1209 & 0.044651 \\
6.4693 & 3.5608e-06 & 0.964203 & 0.015014 & 1.0653 & 0.050912 \\
9.5 & 1.6935e-06 & 0.965654 & 0.018618 & 1.0337 & 0.056695 \\
13.951 & 7.1265e-07 & 0.966254 & 0.022059 & 1.0194 & 0.061712 \\
20.486 & 2.8802e-07 & 0.966325 & 0.025151 & 1.0163 & 0.065896 \\
30.083 & 1.1787e-07 & 0.966109 & 0.027841 & 1.0198 & 0.06933 \\
44.177 & 5.0167e-08 & 0.965755 & 0.030148 & 1.027 & 0.072145 \\
64.873 & 2.2439e-08 & 0.965346 & 0.032122 & 1.0361 & 0.07447 \\
95.264 & 1.0566e-08 & 0.964926 & 0.033818 & 1.046 & 0.07641 \\
\bottomrule
\end{tabular}
\end{minipage}\hfill
\begin{minipage}[t]{0.48\textwidth}
\appendixtablecaption{tab:B32}{$\gamma=10$, $n=10$, $N=80$.}
\begin{tabular}{@{}cccccc@{}}
\toprule
$V_0$ & $\lambda$ & $n_s$ & $r$ & $\Delta\phi/M_{\rm Pl}$ & $H/M$ \\
\midrule
0.94737 & 6.2922e-07 & 0.949296 & 0.0029527 & 1.9427 & 0.022578 \\
1.3912 & 9.8565e-07 & 0.955066 & 0.0042448 & 1.7273 & 0.027071 \\
2.0429 & 1.2406e-06 & 0.960081 & 0.0059978 & 1.556 & 0.032229 \\
3 & 1.1563e-06 & 0.96424 & 0.008324 & 1.4286 & 0.037909 \\
4.4054 & 8.1544e-07 & 0.967262 & 0.011131 & 1.339 & 0.043837 \\
6.4693 & 4.4246e-07 & 0.969173 & 0.014243 & 1.2832 & 0.049588 \\
9.5 & 1.9970e-07 & 0.970153 & 0.017378 & 1.2538 & 0.054774 \\
13.951 & 8.2253e-08 & 0.97049 & 0.0203 & 1.2428 & 0.0592 \\
20.486 & 3.3248e-08 & 0.970446 & 0.02289 & 1.2432 & 0.062864 \\
30.083 & 1.3768e-08 & 0.970204 & 0.025128 & 1.2502 & 0.065865 \\
44.177 & 5.9587e-09 & 0.96987 & 0.027044 & 1.2607 & 0.068331 \\
64.873 & 2.7135e-09 & 0.969505 & 0.028685 & 1.273 & 0.070374 \\
95.264 & 1.3004e-09 & 0.969138 & 0.030098 & 1.286 & 0.072086 \\
\bottomrule
\end{tabular}
\end{minipage}
\par\medskip

\noindent\begin{minipage}[t]{0.48\textwidth}
\appendixtablecaption{tab:B33}{$\gamma=10$, $n=12$, $N=50$.}
\begin{tabular}{@{}cccccc@{}}
\toprule
$V_0$ & $\lambda$ & $n_s$ & $r$ & $\Delta\phi/M_{\rm Pl}$ & $H/M$ \\
\midrule
3 & 2.2004e-05 & 0.936355 & 0.0088841 & 0.7904 & 0.039164 \\
4.4054 & 2.0779e-05 & 0.94283 & 0.012351 & 0.72141 & 0.046177 \\
6.4693 & 1.3887e-05 & 0.947705 & 0.016626 & 0.67192 & 0.053575 \\
9.5 & 6.7397e-06 & 0.950893 & 0.021453 & 0.63962 & 0.060858 \\
13.951 & 2.5863e-06 & 0.952605 & 0.026409 & 0.62113 & 0.067523 \\
20.486 & 8.7615e-07 & 0.953254 & 0.031099 & 0.61247 & 0.073275 \\
30.083 & 2.8730e-07 & 0.953251 & 0.0353 & 0.61012 & 0.078067 \\
44.177 & 9.6531e-08 & 0.952899 & 0.038952 & 0.61152 & 0.082006 \\
64.873 & 3.4170e-08 & 0.952382 & 0.04209 & 0.61502 & 0.085244 \\
95.264 & 1.2871e-08 & 0.951802 & 0.04478 & 0.61965 & 0.087927 \\
\bottomrule
\end{tabular}
\end{minipage}\hfill
\begin{minipage}[t]{0.48\textwidth}
\appendixtablecaption{tab:B34}{$\gamma=10$, $n=12$, $N=60$.}
\begin{tabular}{@{}cccccc@{}}
\toprule
$V_0$ & $\lambda$ & $n_s$ & $r$ & $\Delta\phi/M_{\rm Pl}$ & $H/M$ \\
\midrule
1.3912 & 6.4068e-07 & 0.936534 & 0.004309 & 1.2245 & 0.027275 \\
2.0429 & 9.6686e-07 & 0.943496 & 0.0061704 & 1.0921 & 0.032639 \\
3 & 1.0884e-06 & 0.949518 & 0.0086879 & 0.98836 & 0.038729 \\
4.4054 & 8.7555e-07 & 0.954322 & 0.011914 & 0.91128 & 0.045353 \\
6.4693 & 5.0059e-07 & 0.957727 & 0.015739 & 0.85844 & 0.052128 \\
9.5 & 2.1499e-07 & 0.959775 & 0.01987 & 0.82618 & 0.05857 \\
13.951 & 7.6859e-08 & 0.960742 & 0.023941 & 0.80954 & 0.064291 \\
20.486 & 2.5426e-08 & 0.960989 & 0.027683 & 0.80345 & 0.069133 \\
30.083 & 8.3969e-09 & 0.960824 & 0.03098 & 0.80384 & 0.073134 \\
44.177 & 2.8868e-09 & 0.96045 & 0.033825 & 0.80795 & 0.076418 \\
64.873 & 1.0521e-09 & 0.959984 & 0.036266 & 0.81409 & 0.079127 \\
95.264 & 4.0823e-10 & 0.959491 & 0.038362 & 0.82127 & 0.081379 \\
\bottomrule
\end{tabular}
\end{minipage}
\par\medskip

\noindent\begin{minipage}[t]{0.48\textwidth}
\appendixtablecaption{tab:B35}{$\gamma=10$, $n=12$, $N=70$.}
\begin{tabular}{@{}cccccc@{}}
\toprule
$V_0$ & $\lambda$ & $n_s$ & $r$ & $\Delta\phi/M_{\rm Pl}$ & $H/M$ \\
\midrule
0.94737 & 3.4263e-08 & 0.941607 & 0.0029636 & 1.6525 & 0.02262 \\
1.3912 & 6.0151e-08 & 0.947958 & 0.0042727 & 1.4643 & 0.02716 \\
2.0429 & 8.1354e-08 & 0.953644 & 0.0060835 & 1.3139 & 0.032408 \\
3 & 8.0494e-08 & 0.958425 & 0.0084871 & 1.1982 & 0.038279 \\
4.4054 & 5.6441e-08 & 0.962082 & 0.011482 & 1.115 & 0.044524 \\
6.4693 & 2.8538e-08 & 0.964526 & 0.014908 & 1.0605 & 0.050733 \\
9.5 & 1.1250e-08 & 0.96588 & 0.018471 & 1.0295 & 0.05647 \\
13.951 & 3.8583e-09 & 0.966427 & 0.021875 & 1.0155 & 0.061455 \\
20.486 & 1.2673e-09 & 0.966472 & 0.024943 & 1.0125 & 0.065623 \\
30.083 & 4.2390e-10 & 0.966242 & 0.027619 & 1.016 & 0.069052 \\
44.177 & 1.4919e-10 & 0.96588 & 0.029919 & 1.0231 & 0.071874 \\
64.873 & 5.5762e-11 & 0.965465 & 0.031892 & 1.0322 & 0.074206 \\
95.264 & 2.2096e-11 & 0.96504 & 0.033592 & 1.0424 & 0.076135 \\
\bottomrule
\end{tabular}
\end{minipage}\hfill
\begin{minipage}[t]{0.48\textwidth}
\appendixtablecaption{tab:B36}{$\gamma=10$, $n=12$, $N=80$.}
\begin{tabular}{@{}cccccc@{}}
\toprule
$V_0$ & $\lambda$ & $n_s$ & $r$ & $\Delta\phi/M_{\rm Pl}$ & $H/M$ \\
\midrule
0.94737 & 4.6299e-09 & 0.950808 & 0.002948 & 1.9244 & 0.02256 \\
1.3912 & 7.4774e-09 & 0.956193 & 0.0042349 & 1.7128 & 0.02704 \\
2.0429 & 9.1341e-09 & 0.960925 & 0.0059938 & 1.5456 & 0.032168 \\
3 & 8.0442e-09 & 0.964793 & 0.0082845 & 1.4198 & 0.037819 \\
4.4054 & 5.0165e-09 & 0.967631 & 0.011062 & 1.3321 & 0.043702 \\
6.4693 & 2.3031e-09 & 0.969421 & 0.014137 & 1.2774 & 0.049404 \\
9.5 & 8.5513e-10 & 0.970332 & 0.017237 & 1.2486 & 0.054551 \\
13.951 & 2.8642e-10 & 0.970633 & 0.020131 & 1.2379 & 0.058953 \\
20.486 & 9.4220e-11 & 0.97057 & 0.022702 & 1.2383 & 0.062606 \\
30.083 & 3.2168e-11 & 0.970318 & 0.024927 & 1.2449 & 0.065623 \\
44.177 & 1.1617e-11 & 0.969979 & 0.026836 & 1.255 & 0.068117 \\
64.873 & 4.2595e-12 & 0.969604 & 0.0285 & 1.2704 & 0.070011 \\
95.264 & 1.7380e-12 & 0.969234 & 0.029912 & 1.2833 & 0.071735 \\
\bottomrule
\end{tabular}
\end{minipage}
\par\medskip

\noindent\begin{minipage}[t]{0.48\textwidth}
\appendixtablecaption{tab:B37}{$\gamma=12$, $n=8$, $N=50$.}
\begin{tabular}{@{}cccccc@{}}
\toprule
$V_0$ & $\lambda$ & $n_s$ & $r$ & $\Delta\phi/M_{\rm Pl}$ & $H/M$ \\
\midrule
3 & 0.019089 & 0.934389 & 0.0088714 & 0.79046 & 0.039136 \\
4.4054 & 0.019586 & 0.941743 & 0.012294 & 0.72042 & 0.04607 \\
6.4693 & 0.015732 & 0.947094 & 0.016452 & 0.67109 & 0.053295 \\
9.5 & 0.01 & 0.950469 & 0.021074 & 0.64025 & 0.060251 \\
13.951 & 0.0055502 & 0.952183 & 0.025646 & 0.62238 & 0.06654 \\
20.486 & 0.0028107 & 0.952802 & 0.029896 & 0.61469 & 0.071843 \\
30.083 & 0.0013963 & 0.952778 & 0.033631 & 0.61298 & 0.076199 \\
44.177 & 7.0535e-04 & 0.952423 & 0.036835 & 0.61473 & 0.079746 \\
64.873 & 3.6841e-04 & 0.951919 & 0.039562 & 0.61837 & 0.082645 \\
95.264 & 2.0010e-04 & 0.951361 & 0.041886 & 0.623 & 0.085037 \\
\bottomrule
\end{tabular}
\end{minipage}\hfill
\begin{minipage}[t]{0.48\textwidth}
\appendixtablecaption{tab:B38}{$\gamma=12$, $n=8$, $N=60$.}
\begin{tabular}{@{}cccccc@{}}
\toprule
$V_0$ & $\lambda$ & $n_s$ & $r$ & $\Delta\phi/M_{\rm Pl}$ & $H/M$ \\
\midrule
1.3912 & 0.0015601 & 0.933054 & 0.0043107 & 1.2353 & 0.02728 \\
2.0429 & 0.0022312 & 0.94131 & 0.0061674 & 1.098 & 0.032631 \\
3 & 0.0025937 & 0.948262 & 0.0086649 & 0.99148 & 0.038678 \\
4.4054 & 0.0023748 & 0.953636 & 0.01183 & 0.91351 & 0.045194 \\
6.4693 & 0.0017099 & 0.957306 & 0.015517 & 0.86124 & 0.051758 \\
9.5 & 0.0010114 & 0.959423 & 0.019403 & 0.83029 & 0.057877 \\
13.951 & 5.3000e-04 & 0.960377 & 0.023137 & 0.81501 & 0.063202 \\
20.486 & 2.6454e-04 & 0.960601 & 0.026496 & 0.80996 & 0.067635 \\
30.083 & 1.3210e-04 & 0.960425 & 0.029407 & 0.81098 & 0.071254 \\
44.177 & 6.7725e-05 & 0.960055 & 0.031892 & 0.81541 & 0.074202 \\
64.873 & 3.6030e-05 & 0.959605 & 0.034006 & 0.82163 & 0.076622 \\
95.264 & 1.9942e-05 & 0.959133 & 0.035812 & 0.82872 & 0.078631 \\
\bottomrule
\end{tabular}
\end{minipage}
\par\medskip

\noindent\begin{minipage}[t]{0.48\textwidth}
\appendixtablecaption{tab:B39}{$\gamma=12$, $n=8$, $N=70$.}
\begin{tabular}{@{}cccccc@{}}
\toprule
$V_0$ & $\lambda$ & $n_s$ & $r$ & $\Delta\phi/M_{\rm Pl}$ & $H/M$ \\
\midrule
0.94737 & 2.0556e-04 & 0.9378 & 0.0029653 & 1.6768 & 0.022626 \\
1.3912 & 3.2691e-04 & 0.945445 & 0.0042733 & 1.48 & 0.027162 \\
2.0429 & 4.3277e-04 & 0.952119 & 0.0060765 & 1.324 & 0.03239 \\
3 & 4.5898e-04 & 0.957574 & 0.0084527 & 1.2054 & 0.038201 \\
4.4054 & 3.8133e-04 & 0.961609 & 0.011373 & 1.1216 & 0.044312 \\
6.4693 & 2.5213e-04 & 0.964201 & 0.014645 & 1.0681 & 0.050283 \\
9.5 & 1.4084e-04 & 0.965576 & 0.017963 & 1.0388 & 0.055688 \\
13.951 & 7.1871e-05 & 0.966105 & 0.021058 & 1.0263 & 0.060295 \\
20.486 & 3.5733e-05 & 0.966132 & 0.023792 & 1.0244 & 0.064091 \\
30.083 & 1.7999e-05 & 0.965899 & 0.026143 & 1.0285 & 0.067182 \\
44.177 & 9.3601e-06 & 0.965545 & 0.028144 & 1.036 & 0.069707 \\
64.873 & 5.0593e-06 & 0.965146 & 0.02985 & 1.0451 & 0.071787 \\
95.264 & 2.8447e-06 & 0.96474 & 0.03131 & 1.0549 & 0.073523 \\
\bottomrule
\end{tabular}
\end{minipage}\hfill
\begin{minipage}[t]{0.48\textwidth}
\appendixtablecaption{tab:B40}{$\gamma=12$, $n=8$, $N=80$.}
\begin{tabular}{@{}cccccc@{}}
\toprule
$V_0$ & $\lambda$ & $n_s$ & $r$ & $\Delta\phi/M_{\rm Pl}$ & $H/M$ \\
\midrule
0.94737 & 5.4894e-05 & 0.947891 & 0.0029495 & 1.9555 & 0.022566 \\
1.3912 & 8.2342e-05 & 0.954319 & 0.0042342 & 1.7341 & 0.027037 \\
2.0429 & 1.0142e-04 & 0.959822 & 0.0059823 & 1.5608 & 0.032138 \\
3 & 9.9022e-05 & 0.964188 & 0.0082381 & 1.4322 & 0.037713 \\
4.4054 & 7.5754e-05 & 0.967279 & 0.010929 & 1.3442 & 0.043439 \\
6.4693 & 4.6880e-05 & 0.969151 & 0.013842 & 1.2911 & 0.048885 \\
9.5 & 2.5174e-05 & 0.970061 & 0.016703 & 1.2643 & 0.0537 \\
13.951 & 1.2659e-05 & 0.970343 & 0.019314 & 1.2552 & 0.057745 \\
20.486 & 6.3035e-06 & 0.97027 & 0.021594 & 1.2567 & 0.061058 \\
30.083 & 3.2064e-06 & 0.970018 & 0.023544 & 1.2643 & 0.063756 \\
44.177 & 1.6894e-06 & 0.969688 & 0.025204 & 1.275 & 0.065965 \\
64.873 & 9.2588e-07 & 0.969333 & 0.02662 & 1.2873 & 0.067793 \\
95.264 & 5.2758e-07 & 0.96898 & 0.027836 & 1.3002 & 0.069324 \\
\bottomrule
\end{tabular}
\end{minipage}
\par\medskip

\noindent\begin{minipage}[t]{0.48\textwidth}
\appendixtablecaption{tab:B41}{$\gamma=12$, $n=10$, $N=50$.}
\begin{tabular}{@{}cccccc@{}}
\toprule
$V_0$ & $\lambda$ & $n_s$ & $r$ & $\Delta\phi/M_{\rm Pl}$ & $H/M$ \\
\midrule
3 & 7.7458e-04 & 0.936462 & 0.0088263 & 0.78256 & 0.039036 \\
4.4054 & 7.4310e-04 & 0.943141 & 0.012206 & 0.71443 & 0.045905 \\
6.4693 & 5.2838e-04 & 0.948006 & 0.016298 & 0.66651 & 0.053045 \\
9.5 & 2.8745e-04 & 0.951058 & 0.020811 & 0.63603 & 0.059941 \\
13.951 & 1.2975e-04 & 0.952611 & 0.025331 & 0.61915 & 0.066131 \\
20.486 & 5.3435e-05 & 0.953141 & 0.029521 & 0.61164 & 0.071391 \\
30.083 & 2.1607e-05 & 0.953071 & 0.033218 & 0.60997 & 0.07573 \\
44.177 & 8.9550e-06 & 0.952692 & 0.036403 & 0.61172 & 0.079277 \\
64.873 & 3.8799e-06 & 0.952172 & 0.039123 & 0.61535 & 0.082185 \\
95.264 & 1.7689e-06 & 0.951604 & 0.041449 & 0.61995 & 0.084593 \\
\bottomrule
\end{tabular}
\end{minipage}\hfill
\begin{minipage}[t]{0.48\textwidth}
\appendixtablecaption{tab:B42}{$\gamma=12$, $n=10$, $N=60$.}
\begin{tabular}{@{}cccccc@{}}
\toprule
$V_0$ & $\lambda$ & $n_s$ & $r$ & $\Delta\phi/M_{\rm Pl}$ & $H/M$ \\
\midrule
1.3912 & 3.7948e-05 & 0.935961 & 0.0042986 & 1.2192 & 0.027242 \\
2.0429 & 5.5082e-05 & 0.943409 & 0.0061416 & 1.0855 & 0.032563 \\
3 & 6.1679e-05 & 0.949711 & 0.0086128 & 0.98189 & 0.038561 \\
4.4054 & 5.1446e-05 & 0.954596 & 0.011733 & 0.90617 & 0.045008 \\
6.4693 & 3.2054e-05 & 0.957932 & 0.015357 & 0.85546 & 0.051491 \\
9.5 & 1.5801e-05 & 0.959849 & 0.019173 & 0.82542 & 0.057534 \\
13.951 & 6.7624e-06 & 0.960695 & 0.022848 & 0.81056 & 0.062806 \\
20.486 & 2.7432e-06 & 0.960864 & 0.026165 & 0.80565 & 0.067211 \\
30.083 & 1.1186e-06 & 0.96066 & 0.029052 & 0.8067 & 0.070822 \\
44.177 & 4.7291e-07 & 0.960274 & 0.031526 & 0.81111 & 0.073776 \\
64.873 & 2.0987e-07 & 0.959813 & 0.033639 & 0.81731 & 0.076208 \\
95.264 & 9.8054e-08 & 0.959333 & 0.035449 & 0.82439 & 0.078231 \\
\bottomrule
\end{tabular}
\end{minipage}
\par\medskip

\noindent\begin{minipage}[t]{0.48\textwidth}
\appendixtablecaption{tab:B43}{$\gamma=12$, $n=10$, $N=70$.}
\begin{tabular}{@{}cccccc@{}}
\toprule
$V_0$ & $\lambda$ & $n_s$ & $r$ & $\Delta\phi/M_{\rm Pl}$ & $H/M$ \\
\midrule
0.94737 & 3.1470e-06 & 0.940783 & 0.0029589 & 1.6527 & 0.022602 \\
1.3912 & 5.2229e-06 & 0.947652 & 0.0042593 & 1.4611 & 0.027117 \\
2.0429 & 6.8792e-06 & 0.953685 & 0.006047 & 1.3092 & 0.032311 \\
3 & 6.8751e-06 & 0.958637 & 0.0083947 & 1.1941 & 0.03807 \\
4.4054 & 5.0934e-06 & 0.962305 & 0.01127 & 1.1128 & 0.04411 \\
6.4693 & 2.8655e-06 & 0.964659 & 0.014484 & 1.061 & 0.050006 \\
9.5 & 1.3207e-06 & 0.965898 & 0.017744 & 1.0326 & 0.055348 \\
13.951 & 5.4861e-07 & 0.966355 & 0.020793 & 1.0205 & 0.059915 \\
20.486 & 2.2193e-07 & 0.966347 & 0.023498 & 1.0186 & 0.063693 \\
30.083 & 9.1613e-08 & 0.966095 & 0.025833 & 1.0228 & 0.066783 \\
44.177 & 3.9464e-08 & 0.965729 & 0.027829 & 1.0302 & 0.069315 \\
64.873 & 1.7880e-08 & 0.965322 & 0.029535 & 1.0393 & 0.071408 \\
95.264 & 8.5250e-09 & 0.96491 & 0.031 & 1.0491 & 0.073158 \\
\bottomrule
\end{tabular}
\end{minipage}\hfill
\begin{minipage}[t]{0.48\textwidth}
\appendixtablecaption{tab:B44}{$\gamma=12$, $n=10$, $N=80$.}
\begin{tabular}{@{}cccccc@{}}
\toprule
$V_0$ & $\lambda$ & $n_s$ & $r$ & $\Delta\phi/M_{\rm Pl}$ & $H/M$ \\
\midrule
0.94737 & 5.8912e-07 & 0.950257 & 0.002942 & 1.9277 & 0.022537 \\
1.3912 & 9.0807e-07 & 0.956046 & 0.0042182 & 1.7123 & 0.026986 \\
2.0429 & 1.0929e-06 & 0.96103 & 0.0059494 & 1.5439 & 0.032049 \\
3 & 9.8632e-07 & 0.964996 & 0.0081753 & 1.4191 & 0.037569 \\
4.4054 & 6.6119e-07 & 0.967806 & 0.010822 & 1.3339 & 0.043225 \\
6.4693 & 3.4383e-07 & 0.969503 & 0.013682 & 1.2825 & 0.048602 \\
9.5 & 1.5137e-07 & 0.970317 & 0.016496 & 1.2565 & 0.053366 \\
13.951 & 6.1880e-08 & 0.97055 & 0.019072 & 1.2477 & 0.057382 \\
20.486 & 2.5119e-08 & 0.970451 & 0.02133 & 1.2493 & 0.060684 \\
30.083 & 1.0508e-08 & 0.970186 & 0.02327 & 1.2569 & 0.063384 \\
44.177 & 4.6052e-09 & 0.969847 & 0.024927 & 1.2676 & 0.065602 \\
64.873 & 2.1242e-09 & 0.969486 & 0.026345 & 1.2799 & 0.067442 \\
95.264 & 1.0304e-09 & 0.969128 & 0.027566 & 1.2927 & 0.068987 \\
\bottomrule
\end{tabular}
\end{minipage}
\par\medskip

\noindent\begin{minipage}[t]{0.48\textwidth}
\appendixtablecaption{tab:B45}{$\gamma=12$, $n=12$, $N=50$.}
\begin{tabular}{@{}cccccc@{}}
\toprule
$V_0$ & $\lambda$ & $n_s$ & $r$ & $\Delta\phi/M_{\rm Pl}$ & $H/M$ \\
\midrule
2.0429 & 1.6707e-05 & 0.930024 & 0.0062176 & 0.86633 & 0.032764 \\
3 & 2.0974e-05 & 0.937739 & 0.0087962 & 0.77758 & 0.03897 \\
4.4054 & 1.8820e-05 & 0.94401 & 0.012148 & 0.71064 & 0.045795 \\
6.4693 & 1.1852e-05 & 0.948581 & 0.016197 & 0.66359 & 0.05288 \\
9.5 & 5.4593e-06 & 0.951443 & 0.020657 & 0.63367 & 0.059719 \\
13.951 & 2.0276e-06 & 0.952888 & 0.025126 & 0.61707 & 0.065863 \\
20.486 & 6.7913e-07 & 0.953361 & 0.029277 & 0.60966 & 0.071095 \\
30.083 & 2.2354e-07 & 0.953263 & 0.032949 & 0.60803 & 0.075423 \\
44.177 & 7.6008e-08 & 0.952868 & 0.036121 & 0.60976 & 0.078969 \\
64.873 & 2.7317e-08 & 0.952339 & 0.038836 & 0.61338 & 0.081884 \\
95.264 & 1.0454e-08 & 0.951764 & 0.041163 & 0.61797 & 0.084301 \\
\bottomrule
\end{tabular}
\end{minipage}\hfill
\begin{minipage}[t]{0.48\textwidth}
\appendixtablecaption{tab:B46}{$\gamma=12$, $n=12$, $N=60$.}
\begin{tabular}{@{}cccccc@{}}
\toprule
$V_0$ & $\lambda$ & $n_s$ & $r$ & $\Delta\phi/M_{\rm Pl}$ & $H/M$ \\
\midrule
1.3912 & 6.1609e-07 & 0.93773 & 0.0042904 & 1.2091 & 0.027216 \\
2.0429 & 9.0736e-07 & 0.944696 & 0.0061244 & 1.0776 & 0.032517 \\
3 & 9.7890e-07 & 0.950608 & 0.0085781 & 0.97584 & 0.038484 \\
4.4054 & 7.4414e-07 & 0.955196 & 0.01167 & 0.90152 & 0.044885 \\
6.4693 & 4.0145e-07 & 0.958329 & 0.015252 & 0.85177 & 0.051315 \\
9.5 & 1.6498e-07 & 0.960122 & 0.019024 & 0.82229 & 0.05731 \\
13.951 & 5.7680e-08 & 0.9609 & 0.02266 & 0.80769 & 0.062548 \\
20.486 & 1.9017e-08 & 0.961035 & 0.02595 & 0.80286 & 0.066934 \\
30.083 & 6.3329e-09 & 0.960814 & 0.028821 & 0.80392 & 0.07054 \\
44.177 & 2.2076e-09 & 0.960418 & 0.031288 & 0.80831 & 0.073496 \\
64.873 & 8.1718e-10 & 0.95995 & 0.033399 & 0.81451 & 0.075935 \\
95.264 & 3.2250e-10 & 0.959465 & 0.035211 & 0.82154 & 0.077971 \\
\bottomrule
\end{tabular}
\end{minipage}
\par\medskip

\noindent\begin{minipage}[t]{0.48\textwidth}
\appendixtablecaption{tab:B47}{$\gamma=12$, $n=12$, $N=70$.}
\begin{tabular}{@{}cccccc@{}}
\toprule
$V_0$ & $\lambda$ & $n_s$ & $r$ & $\Delta\phi/M_{\rm Pl}$ & $H/M$ \\
\midrule
0.94737 & 3.2162e-08 & 0.942589 & 0.0029545 & 1.6376 & 0.022585 \\
1.3912 & 5.5684e-08 & 0.948998 & 0.0042498 & 1.4492 & 0.027087 \\
2.0429 & 7.2968e-08 & 0.954649 & 0.0060274 & 1.3 & 0.032258 \\
3 & 6.8741e-08 & 0.959297 & 0.0083563 & 1.187 & 0.037983 \\
4.4054 & 4.5436e-08 & 0.962743 & 0.011202 & 1.1072 & 0.043977 \\
6.4693 & 2.1762e-08 & 0.964951 & 0.014379 & 1.0565 & 0.049824 \\
9.5 & 8.2783e-09 & 0.966105 & 0.017601 & 1.0286 & 0.055125 \\
13.951 & 2.7995e-09 & 0.966518 & 0.020621 & 1.0167 & 0.059667 \\
20.486 & 9.2145e-10 & 0.966488 & 0.023306 & 1.0149 & 0.063433 \\
30.083 & 3.1156e-10 & 0.966223 & 0.025631 & 1.0191 & 0.06652 \\
44.177 & 1.1134e-10 & 0.96585 & 0.027622 & 1.0264 & 0.069061 \\
64.873 & 4.2280e-11 & 0.965438 & 0.029329 & 1.0354 & 0.071162 \\
95.264 & 1.6917e-11 & 0.96502 & 0.030802 & 1.0459 & 0.072878 \\
\bottomrule
\end{tabular}
\end{minipage}\hfill
\begin{minipage}[t]{0.48\textwidth}
\appendixtablecaption{tab:B48}{$\gamma=12$, $n=12$, $N=80$.}
\begin{tabular}{@{}cccccc@{}}
\toprule
$V_0$ & $\lambda$ & $n_s$ & $r$ & $\Delta\phi/M_{\rm Pl}$ & $H/M$ \\
\midrule
0.94737 & 4.1521e-09 & 0.951755 & 0.0029485 & 1.9113 & 0.022516 \\
1.3912 & 6.6824e-09 & 0.957103 & 0.0042075 & 1.6985 & 0.026952 \\
2.0429 & 7.8589e-09 & 0.961775 & 0.0059275 & 1.5332 & 0.03199 \\
3 & 6.5589e-09 & 0.965499 & 0.0081338 & 1.4108 & 0.037474 \\
4.4054 & 3.8550e-09 & 0.968138 & 0.010752 & 1.3273 & 0.043085 \\
6.4693 & 1.6853e-09 & 0.969728 & 0.013578 & 1.2769 & 0.048418 \\
9.5 & 6.0845e-10 & 0.970483 & 0.016361 & 1.2515 & 0.053148 \\
13.951 & 2.0224e-10 & 0.970685 & 0.018914 & 1.2429 & 0.057144 \\
20.486 & 6.6843e-11 & 0.97057 & 0.021159 & 1.2446 & 0.060436 \\
30.083 & 2.3159e-11 & 0.970296 & 0.023088 & 1.2517 & 0.063158 \\
44.177 & 8.6030e-12 & 0.969955 & 0.024734 & 1.2609 & 0.065448 \\
64.873 & 3.1540e-12 & 0.969582 & 0.02618 & 1.2775 & 0.067092 \\
\bottomrule
\end{tabular}
\end{minipage}
\par\medskip

\noindent\begin{minipage}[t]{\textwidth}
\centering
\appendixtablecaption{tab:B49}{Evolution along the $\gamma$ branch for $n=10$, $M=2.45\times10^{-4}$, $V_0=4.4054\times10^{-10}$, $N=80$.}
\begin{tabular}{@{}ccccccc@{}}
\toprule
$\gamma$ & $c_s$ & $\lambda$ & $n_s$ & $r$ & $\Delta\phi/M_{\rm Pl}$ & $H/M$ \\
\midrule
3 & 0.44721 & 5.5803e-08 & 0.964045 & 0.0125433 & 2.01119 & 0.0465355 \\
4 & 0.37796 & 2.6650e-07 & 0.964478 & 0.0123076 & 1.67383 & 0.0460962 \\
5 & 0.33333 & 5.4666e-07 & 0.965061 & 0.0120806 & 1.52043 & 0.0456692 \\
6 & 0.30151 & 7.6535e-07 & 0.96562 & 0.0118665 & 1.43894 & 0.0452627 \\
8 & 0.2582 & 9.0768e-07 & 0.966554 & 0.0114765 & 1.36443 & 0.0445126 \\
10 & 0.22942 & 8.1544e-07 & 0.967262 & 0.011131 & 1.33904 & 0.0438374 \\
12 & 0.20851 & 6.6119e-07 & 0.967806 & 0.0108222 & 1.33393 & 0.0432251 \\
16 & 0.17961 & 3.9818e-07 & 0.968572 & 0.0102909 & 1.34831 & 0.0421507 \\
20 & 0.16013 & 2.3905e-07 & 0.969078 & 0.00984653 & 1.375 & 0.0412306 \\
30 & 0.13019 & 7.8288e-08 & 0.969798 & 0.00898546 & 1.45193 & 0.0393866 \\
40 & 0.11251 & 3.1914e-08 & 0.970164 & 0.00834899 & 1.52576 & 0.037966 \\
50 & 0.1005 & 1.5262e-08 & 0.970375 & 0.0078503 & 1.59299 & 0.0368147 \\
\bottomrule
\end{tabular}
\end{minipage}
\par\medskip
\endgroup

\bibliographystyle{cpc}
\bibliography{references}

\end{document}